\documentclass[useAMS,usenatbib,usegraphicx]{mn2e}
\usepackage{epsfig}
\usepackage{color} 
  \definecolor{snow}{rgb}{1.000000,0.980392,0.980392}
  \definecolor{ghost white}{rgb}{0.972549,0.972549,1.000000}
  \definecolor{GhostWhite}{rgb}{0.972549,0.972549,1.000000}
  \definecolor{white smoke}{rgb}{0.960784,0.960784,0.960784}
  \definecolor{WhiteSmoke}{rgb}{0.960784,0.960784,0.960784}
  \definecolor{gainsboro}{rgb}{0.862745,0.862745,0.862745}
  \definecolor{floral white}{rgb}{1.000000,0.980392,0.941176}
  \definecolor{FloralWhite}{rgb}{1.000000,0.980392,0.941176}
  \definecolor{old lace}{rgb}{0.992157,0.960784,0.901961}
  \definecolor{OldLace}{rgb}{0.992157,0.960784,0.901961}
  \definecolor{linen}{rgb}{0.980392,0.941176,0.901961}
  \definecolor{antique white}{rgb}{0.980392,0.921569,0.843137}
  \definecolor{AntiqueWhite}{rgb}{0.980392,0.921569,0.843137}
  \definecolor{papaya whip}{rgb}{1.000000,0.937255,0.835294}
  \definecolor{PapayaWhip}{rgb}{1.000000,0.937255,0.835294}
  \definecolor{blanched almond}{rgb}{1.000000,0.921569,0.803922}
  \definecolor{BlanchedAlmond}{rgb}{1.000000,0.921569,0.803922}
  \definecolor{bisque}{rgb}{1.000000,0.894118,0.768627}
  \definecolor{peach puff}{rgb}{1.000000,0.854902,0.725490}
  \definecolor{PeachPuff}{rgb}{1.000000,0.854902,0.725490}
  \definecolor{navajo white}{rgb}{1.000000,0.870588,0.678431}
  \definecolor{NavajoWhite}{rgb}{1.000000,0.870588,0.678431}
  \definecolor{moccasin}{rgb}{1.000000,0.894118,0.709804}
  \definecolor{cornsilk}{rgb}{1.000000,0.972549,0.862745}
  \definecolor{ivory}{rgb}{1.000000,1.000000,0.941176}
  \definecolor{lemon chiffon}{rgb}{1.000000,0.980392,0.803922}
  \definecolor{LemonChiffon}{rgb}{1.000000,0.980392,0.803922}
  \definecolor{seashell}{rgb}{1.000000,0.960784,0.933333}
  \definecolor{honeydew}{rgb}{0.941176,1.000000,0.941176}
  \definecolor{mint cream}{rgb}{0.960784,1.000000,0.980392}
  \definecolor{MintCream}{rgb}{0.960784,1.000000,0.980392}
  \definecolor{azure}{rgb}{0.941176,1.000000,1.000000}
  \definecolor{alice blue}{rgb}{0.941176,0.972549,1.000000}
  \definecolor{AliceBlue}{rgb}{0.941176,0.972549,1.000000}
  \definecolor{lavender}{rgb}{0.901961,0.901961,0.980392}
  \definecolor{lavender blush}{rgb}{1.000000,0.941176,0.960784}
  \definecolor{LavenderBlush}{rgb}{1.000000,0.941176,0.960784}
  \definecolor{misty rose}{rgb}{1.000000,0.894118,0.882353}
  \definecolor{MistyRose}{rgb}{1.000000,0.894118,0.882353}
  \definecolor{white}{rgb}{1.000000,1.000000,1.000000}
  \definecolor{black}{rgb}{0.000000,0.000000,0.000000}
  \definecolor{dark slate gray}{rgb}{0.184314,0.309804,0.309804}
  \definecolor{DarkSlateGray}{rgb}{0.184314,0.309804,0.309804}
  \definecolor{dark slate grey}{rgb}{0.184314,0.309804,0.309804}
  \definecolor{DarkSlateGrey}{rgb}{0.184314,0.309804,0.309804}
  \definecolor{dim gray}{rgb}{0.411765,0.411765,0.411765}
  \definecolor{DimGray}{rgb}{0.411765,0.411765,0.411765}
  \definecolor{dim grey}{rgb}{0.411765,0.411765,0.411765}
  \definecolor{DimGrey}{rgb}{0.411765,0.411765,0.411765}
  \definecolor{slate gray}{rgb}{0.439216,0.501961,0.564706}
  \definecolor{SlateGray}{rgb}{0.439216,0.501961,0.564706}
  \definecolor{slate grey}{rgb}{0.439216,0.501961,0.564706}
  \definecolor{SlateGrey}{rgb}{0.439216,0.501961,0.564706}
  \definecolor{light slate gray}{rgb}{0.466667,0.533333,0.600000}
  \definecolor{LightSlateGray}{rgb}{0.466667,0.533333,0.600000}
  \definecolor{light slate grey}{rgb}{0.466667,0.533333,0.600000}
  \definecolor{LightSlateGrey}{rgb}{0.466667,0.533333,0.600000}
  \definecolor{gray}{rgb}{0.745098,0.745098,0.745098}
  \definecolor{grey}{rgb}{0.745098,0.745098,0.745098}
  \definecolor{light grey}{rgb}{0.827451,0.827451,0.827451}
  \definecolor{LightGrey}{rgb}{0.827451,0.827451,0.827451}
  \definecolor{light gray}{rgb}{0.827451,0.827451,0.827451}
  \definecolor{LightGray}{rgb}{0.827451,0.827451,0.827451}
  \definecolor{midnight blue}{rgb}{0.098039,0.098039,0.439216}
  \definecolor{MidnightBlue}{rgb}{0.098039,0.098039,0.439216}
  \definecolor{navy}{rgb}{0.000000,0.000000,0.501961}
  \definecolor{navy blue}{rgb}{0.000000,0.000000,0.501961}
  \definecolor{NavyBlue}{rgb}{0.000000,0.000000,0.501961}
  \definecolor{cornflower blue}{rgb}{0.392157,0.584314,0.929412}
  \definecolor{CornflowerBlue}{rgb}{0.392157,0.584314,0.929412}
  \definecolor{dark slate blue}{rgb}{0.282353,0.239216,0.545098}
  \definecolor{DarkSlateBlue}{rgb}{0.282353,0.239216,0.545098}
  \definecolor{slate blue}{rgb}{0.415686,0.352941,0.803922}
  \definecolor{SlateBlue}{rgb}{0.415686,0.352941,0.803922}
  \definecolor{medium slate blue}{rgb}{0.482353,0.407843,0.933333}
  \definecolor{MediumSlateBlue}{rgb}{0.482353,0.407843,0.933333}
  \definecolor{light slate blue}{rgb}{0.517647,0.439216,1.000000}
  \definecolor{LightSlateBlue}{rgb}{0.517647,0.439216,1.000000}
  \definecolor{medium blue}{rgb}{0.000000,0.000000,0.803922}
  \definecolor{MediumBlue}{rgb}{0.000000,0.000000,0.803922}
  \definecolor{royal blue}{rgb}{0.254902,0.411765,0.882353}
  \definecolor{RoyalBlue}{rgb}{0.254902,0.411765,0.882353}
  \definecolor{blue}{rgb}{0.000000,0.000000,1.000000}
  \definecolor{dodger blue}{rgb}{0.117647,0.564706,1.000000}
  \definecolor{DodgerBlue}{rgb}{0.117647,0.564706,1.000000}
  \definecolor{deep sky blue}{rgb}{0.000000,0.749020,1.000000}
  \definecolor{DeepSkyBlue}{rgb}{0.000000,0.749020,1.000000}
  \definecolor{sky blue}{rgb}{0.529412,0.807843,0.921569}
  \definecolor{SkyBlue}{rgb}{0.529412,0.807843,0.921569}
  \definecolor{light sky blue}{rgb}{0.529412,0.807843,0.980392}
  \definecolor{LightSkyBlue}{rgb}{0.529412,0.807843,0.980392}
  \definecolor{steel blue}{rgb}{0.274510,0.509804,0.705882}
  \definecolor{SteelBlue}{rgb}{0.274510,0.509804,0.705882}
  \definecolor{light steel blue}{rgb}{0.690196,0.768627,0.870588}
  \definecolor{LightSteelBlue}{rgb}{0.690196,0.768627,0.870588}
  \definecolor{light blue}{rgb}{0.678431,0.847059,0.901961}
  \definecolor{LightBlue}{rgb}{0.678431,0.847059,0.901961}
  \definecolor{powder blue}{rgb}{0.690196,0.878431,0.901961}
  \definecolor{PowderBlue}{rgb}{0.690196,0.878431,0.901961}
  \definecolor{pale turquoise}{rgb}{0.686275,0.933333,0.933333}
  \definecolor{PaleTurquoise}{rgb}{0.686275,0.933333,0.933333}
  \definecolor{dark turquoise}{rgb}{0.000000,0.807843,0.819608}
  \definecolor{DarkTurquoise}{rgb}{0.000000,0.807843,0.819608}
  \definecolor{medium turquoise}{rgb}{0.282353,0.819608,0.800000}
  \definecolor{MediumTurquoise}{rgb}{0.282353,0.819608,0.800000}
  \definecolor{turquoise}{rgb}{0.250980,0.878431,0.815686}
  \definecolor{cyan}{rgb}{0.000000,1.000000,1.000000}
  \definecolor{light cyan}{rgb}{0.878431,1.000000,1.000000}
  \definecolor{LightCyan}{rgb}{0.878431,1.000000,1.000000}
  \definecolor{cadet blue}{rgb}{0.372549,0.619608,0.627451}
  \definecolor{CadetBlue}{rgb}{0.372549,0.619608,0.627451}
  \definecolor{medium aquamarine}{rgb}{0.400000,0.803922,0.666667}
  \definecolor{MediumAquamarine}{rgb}{0.400000,0.803922,0.666667}
  \definecolor{aquamarine}{rgb}{0.498039,1.000000,0.831373}
  \definecolor{dark green}{rgb}{0.000000,0.392157,0.000000}
  \definecolor{DarkGreen}{rgb}{0.000000,0.392157,0.000000}
  \definecolor{dark olive green}{rgb}{0.333333,0.419608,0.184314}
  \definecolor{DarkOliveGreen}{rgb}{0.333333,0.419608,0.184314}
  \definecolor{dark sea green}{rgb}{0.560784,0.737255,0.560784}
  \definecolor{DarkSeaGreen}{rgb}{0.560784,0.737255,0.560784}
  \definecolor{sea green}{rgb}{0.180392,0.545098,0.341176}
  \definecolor{SeaGreen}{rgb}{0.180392,0.545098,0.341176}
  \definecolor{medium sea green}{rgb}{0.235294,0.701961,0.443137}
  \definecolor{MediumSeaGreen}{rgb}{0.235294,0.701961,0.443137}
  \definecolor{light sea green}{rgb}{0.125490,0.698039,0.666667}
  \definecolor{LightSeaGreen}{rgb}{0.125490,0.698039,0.666667}
  \definecolor{pale green}{rgb}{0.596078,0.984314,0.596078}
  \definecolor{PaleGreen}{rgb}{0.596078,0.984314,0.596078}
  \definecolor{spring green}{rgb}{0.000000,1.000000,0.498039}
  \definecolor{SpringGreen}{rgb}{0.000000,1.000000,0.498039}
  \definecolor{lawn green}{rgb}{0.486275,0.988235,0.000000}
  \definecolor{LawnGreen}{rgb}{0.486275,0.988235,0.000000}
  \definecolor{green}{rgb}{0.000000,1.000000,0.000000}
  \definecolor{chartreuse}{rgb}{0.498039,1.000000,0.000000}
  \definecolor{medium spring green}{rgb}{0.000000,0.980392,0.603922}
  \definecolor{MediumSpringGreen}{rgb}{0.000000,0.980392,0.603922}
  \definecolor{green yellow}{rgb}{0.678431,1.000000,0.184314}
  \definecolor{GreenYellow}{rgb}{0.678431,1.000000,0.184314}
  \definecolor{lime green}{rgb}{0.196078,0.803922,0.196078}
  \definecolor{LimeGreen}{rgb}{0.196078,0.803922,0.196078}
  \definecolor{yellow green}{rgb}{0.603922,0.803922,0.196078}
  \definecolor{YellowGreen}{rgb}{0.603922,0.803922,0.196078}
  \definecolor{forest green}{rgb}{0.133333,0.545098,0.133333}
  \definecolor{ForestGreen}{rgb}{0.133333,0.545098,0.133333}
  \definecolor{olive drab}{rgb}{0.419608,0.556863,0.137255}
  \definecolor{OliveDrab}{rgb}{0.419608,0.556863,0.137255}
  \definecolor{dark khaki}{rgb}{0.741176,0.717647,0.419608}
  \definecolor{DarkKhaki}{rgb}{0.741176,0.717647,0.419608}
  \definecolor{khaki}{rgb}{0.941176,0.901961,0.549020}
  \definecolor{pale goldenrod}{rgb}{0.933333,0.909804,0.666667}
  \definecolor{PaleGoldenrod}{rgb}{0.933333,0.909804,0.666667}
  \definecolor{light goldenrod yellow}{rgb}{0.980392,0.980392,0.823529}
  \definecolor{LightGoldenrodYellow}{rgb}{0.980392,0.980392,0.823529}
  \definecolor{light yellow}{rgb}{1.000000,1.000000,0.878431}
  \definecolor{LightYellow}{rgb}{1.000000,1.000000,0.878431}
  \definecolor{yellow}{rgb}{1.000000,1.000000,0.000000}
  \definecolor{gold}{rgb}{1.000000,0.843137,0.000000}
  \definecolor{light goldenrod}{rgb}{0.933333,0.866667,0.509804}
  \definecolor{LightGoldenrod}{rgb}{0.933333,0.866667,0.509804}
  \definecolor{goldenrod}{rgb}{0.854902,0.647059,0.125490}
  \definecolor{dark goldenrod}{rgb}{0.721569,0.525490,0.043137}
  \definecolor{DarkGoldenrod}{rgb}{0.721569,0.525490,0.043137}
  \definecolor{rosy brown}{rgb}{0.737255,0.560784,0.560784}
  \definecolor{RosyBrown}{rgb}{0.737255,0.560784,0.560784}
  \definecolor{indian red}{rgb}{0.803922,0.360784,0.360784}
  \definecolor{IndianRed}{rgb}{0.803922,0.360784,0.360784}
  \definecolor{saddle brown}{rgb}{0.545098,0.270588,0.074510}
  \definecolor{SaddleBrown}{rgb}{0.545098,0.270588,0.074510}
  \definecolor{sienna}{rgb}{0.627451,0.321569,0.176471}
  \definecolor{peru}{rgb}{0.803922,0.521569,0.247059}
  \definecolor{burlywood}{rgb}{0.870588,0.721569,0.529412}
  \definecolor{beige}{rgb}{0.960784,0.960784,0.862745}
  \definecolor{wheat}{rgb}{0.960784,0.870588,0.701961}
  \definecolor{sandy brown}{rgb}{0.956863,0.643137,0.376471}
  \definecolor{SandyBrown}{rgb}{0.956863,0.643137,0.376471}
  \definecolor{tan}{rgb}{0.823529,0.705882,0.549020}
  \definecolor{chocolate}{rgb}{0.823529,0.411765,0.117647}
  \definecolor{firebrick}{rgb}{0.698039,0.133333,0.133333}
  \definecolor{brown}{rgb}{0.647059,0.164706,0.164706}
  \definecolor{dark salmon}{rgb}{0.913725,0.588235,0.478431}
  \definecolor{DarkSalmon}{rgb}{0.913725,0.588235,0.478431}
  \definecolor{salmon}{rgb}{0.980392,0.501961,0.447059}
  \definecolor{light salmon}{rgb}{1.000000,0.627451,0.478431}
  \definecolor{LightSalmon}{rgb}{1.000000,0.627451,0.478431}
  \definecolor{orange}{rgb}{1.000000,0.647059,0.000000}
  \definecolor{dark orange}{rgb}{1.000000,0.549020,0.000000}
  \definecolor{DarkOrange}{rgb}{1.000000,0.549020,0.000000}
  \definecolor{coral}{rgb}{1.000000,0.498039,0.313726}
  \definecolor{light coral}{rgb}{0.941176,0.501961,0.501961}
  \definecolor{LightCoral}{rgb}{0.941176,0.501961,0.501961}
  \definecolor{tomato}{rgb}{1.000000,0.388235,0.278431}
  \definecolor{orange red}{rgb}{1.000000,0.270588,0.000000}
  \definecolor{OrangeRed}{rgb}{1.000000,0.270588,0.000000}
  \definecolor{red}{rgb}{1.000000,0.000000,0.000000}
  \definecolor{hot pink}{rgb}{1.000000,0.411765,0.705882}
  \definecolor{HotPink}{rgb}{1.000000,0.411765,0.705882}
  \definecolor{deep pink}{rgb}{1.000000,0.078431,0.576471}
  \definecolor{DeepPink}{rgb}{1.000000,0.078431,0.576471}
  \definecolor{pink}{rgb}{1.000000,0.752941,0.796078}
  \definecolor{light pink}{rgb}{1.000000,0.713726,0.756863}
  \definecolor{LightPink}{rgb}{1.000000,0.713726,0.756863}
  \definecolor{pale violet red}{rgb}{0.858824,0.439216,0.576471}
  \definecolor{PaleVioletRed}{rgb}{0.858824,0.439216,0.576471}
  \definecolor{maroon}{rgb}{0.690196,0.188235,0.376471}
  \definecolor{medium violet red}{rgb}{0.780392,0.082353,0.521569}
  \definecolor{MediumVioletRed}{rgb}{0.780392,0.082353,0.521569}
  \definecolor{violet red}{rgb}{0.815686,0.125490,0.564706}
  \definecolor{VioletRed}{rgb}{0.815686,0.125490,0.564706}
  \definecolor{magenta}{rgb}{1.000000,0.000000,1.000000}
  \definecolor{violet}{rgb}{0.933333,0.509804,0.933333}
  \definecolor{plum}{rgb}{0.866667,0.627451,0.866667}
  \definecolor{orchid}{rgb}{0.854902,0.439216,0.839216}
  \definecolor{medium orchid}{rgb}{0.729412,0.333333,0.827451}
  \definecolor{MediumOrchid}{rgb}{0.729412,0.333333,0.827451}
  \definecolor{dark orchid}{rgb}{0.600000,0.196078,0.800000}
  \definecolor{DarkOrchid}{rgb}{0.600000,0.196078,0.800000}
  \definecolor{dark violet}{rgb}{0.580392,0.000000,0.827451}
  \definecolor{DarkViolet}{rgb}{0.580392,0.000000,0.827451}
  \definecolor{blue violet}{rgb}{0.541176,0.168627,0.886275}
  \definecolor{BlueViolet}{rgb}{0.541176,0.168627,0.886275}
  \definecolor{purple}{rgb}{0.627451,0.125490,0.941176}
  \definecolor{medium purple}{rgb}{0.576471,0.439216,0.858824}
  \definecolor{MediumPurple}{rgb}{0.576471,0.439216,0.858824}
  \definecolor{thistle}{rgb}{0.847059,0.749020,0.847059}
  \definecolor{snow1}{rgb}{1.000000,0.980392,0.980392}
  \definecolor{snow2}{rgb}{0.933333,0.913725,0.913725}
  \definecolor{snow3}{rgb}{0.803922,0.788235,0.788235}
  \definecolor{snow4}{rgb}{0.545098,0.537255,0.537255}
  \definecolor{seashell1}{rgb}{1.000000,0.960784,0.933333}
  \definecolor{seashell2}{rgb}{0.933333,0.898039,0.870588}
  \definecolor{seashell3}{rgb}{0.803922,0.772549,0.749020}
  \definecolor{seashell4}{rgb}{0.545098,0.525490,0.509804}
  \definecolor{AntiqueWhite1}{rgb}{1.000000,0.937255,0.858824}
  \definecolor{AntiqueWhite2}{rgb}{0.933333,0.874510,0.800000}
  \definecolor{AntiqueWhite3}{rgb}{0.803922,0.752941,0.690196}
  \definecolor{AntiqueWhite4}{rgb}{0.545098,0.513726,0.470588}
  \definecolor{bisque1}{rgb}{1.000000,0.894118,0.768627}
  \definecolor{bisque2}{rgb}{0.933333,0.835294,0.717647}
  \definecolor{bisque3}{rgb}{0.803922,0.717647,0.619608}
  \definecolor{bisque4}{rgb}{0.545098,0.490196,0.419608}
  \definecolor{PeachPuff1}{rgb}{1.000000,0.854902,0.725490}
  \definecolor{PeachPuff2}{rgb}{0.933333,0.796078,0.678431}
  \definecolor{PeachPuff3}{rgb}{0.803922,0.686275,0.584314}
  \definecolor{PeachPuff4}{rgb}{0.545098,0.466667,0.396078}
  \definecolor{NavajoWhite1}{rgb}{1.000000,0.870588,0.678431}
  \definecolor{NavajoWhite2}{rgb}{0.933333,0.811765,0.631373}
  \definecolor{NavajoWhite3}{rgb}{0.803922,0.701961,0.545098}
  \definecolor{NavajoWhite4}{rgb}{0.545098,0.474510,0.368627}
  \definecolor{LemonChiffon1}{rgb}{1.000000,0.980392,0.803922}
  \definecolor{LemonChiffon2}{rgb}{0.933333,0.913725,0.749020}
  \definecolor{LemonChiffon3}{rgb}{0.803922,0.788235,0.647059}
  \definecolor{LemonChiffon4}{rgb}{0.545098,0.537255,0.439216}
  \definecolor{cornsilk1}{rgb}{1.000000,0.972549,0.862745}
  \definecolor{cornsilk2}{rgb}{0.933333,0.909804,0.803922}
  \definecolor{cornsilk3}{rgb}{0.803922,0.784314,0.694118}
  \definecolor{cornsilk4}{rgb}{0.545098,0.533333,0.470588}
  \definecolor{ivory1}{rgb}{1.000000,1.000000,0.941176}
  \definecolor{ivory2}{rgb}{0.933333,0.933333,0.878431}
  \definecolor{ivory3}{rgb}{0.803922,0.803922,0.756863}
  \definecolor{ivory4}{rgb}{0.545098,0.545098,0.513726}
  \definecolor{honeydew1}{rgb}{0.941176,1.000000,0.941176}
  \definecolor{honeydew2}{rgb}{0.878431,0.933333,0.878431}
  \definecolor{honeydew3}{rgb}{0.756863,0.803922,0.756863}
  \definecolor{honeydew4}{rgb}{0.513726,0.545098,0.513726}
  \definecolor{LavenderBlush1}{rgb}{1.000000,0.941176,0.960784}
  \definecolor{LavenderBlush2}{rgb}{0.933333,0.878431,0.898039}
  \definecolor{LavenderBlush3}{rgb}{0.803922,0.756863,0.772549}
  \definecolor{LavenderBlush4}{rgb}{0.545098,0.513726,0.525490}
  \definecolor{MistyRose1}{rgb}{1.000000,0.894118,0.882353}
  \definecolor{MistyRose2}{rgb}{0.933333,0.835294,0.823529}
  \definecolor{MistyRose3}{rgb}{0.803922,0.717647,0.709804}
  \definecolor{MistyRose4}{rgb}{0.545098,0.490196,0.482353}
  \definecolor{azure1}{rgb}{0.941176,1.000000,1.000000}
  \definecolor{azure2}{rgb}{0.878431,0.933333,0.933333}
  \definecolor{azure3}{rgb}{0.756863,0.803922,0.803922}
  \definecolor{azure4}{rgb}{0.513726,0.545098,0.545098}
  \definecolor{SlateBlue1}{rgb}{0.513726,0.435294,1.000000}
  \definecolor{SlateBlue2}{rgb}{0.478431,0.403922,0.933333}
  \definecolor{SlateBlue3}{rgb}{0.411765,0.349020,0.803922}
  \definecolor{SlateBlue4}{rgb}{0.278431,0.235294,0.545098}
  \definecolor{RoyalBlue1}{rgb}{0.282353,0.462745,1.000000}
  \definecolor{RoyalBlue2}{rgb}{0.262745,0.431373,0.933333}
  \definecolor{RoyalBlue3}{rgb}{0.227451,0.372549,0.803922}
  \definecolor{RoyalBlue4}{rgb}{0.152941,0.250980,0.545098}
  \definecolor{blue1}{rgb}{0.000000,0.000000,1.000000}
  \definecolor{blue2}{rgb}{0.000000,0.000000,0.933333}
  \definecolor{blue3}{rgb}{0.000000,0.000000,0.803922}
  \definecolor{blue4}{rgb}{0.000000,0.000000,0.545098}
  \definecolor{DodgerBlue1}{rgb}{0.117647,0.564706,1.000000}
  \definecolor{DodgerBlue2}{rgb}{0.109804,0.525490,0.933333}
  \definecolor{DodgerBlue3}{rgb}{0.094118,0.454902,0.803922}
  \definecolor{DodgerBlue4}{rgb}{0.062745,0.305882,0.545098}
  \definecolor{SteelBlue1}{rgb}{0.388235,0.721569,1.000000}
  \definecolor{SteelBlue2}{rgb}{0.360784,0.674510,0.933333}
  \definecolor{SteelBlue3}{rgb}{0.309804,0.580392,0.803922}
  \definecolor{SteelBlue4}{rgb}{0.211765,0.392157,0.545098}
  \definecolor{DeepSkyBlue1}{rgb}{0.000000,0.749020,1.000000}
  \definecolor{DeepSkyBlue2}{rgb}{0.000000,0.698039,0.933333}
  \definecolor{DeepSkyBlue3}{rgb}{0.000000,0.603922,0.803922}
  \definecolor{DeepSkyBlue4}{rgb}{0.000000,0.407843,0.545098}
  \definecolor{SkyBlue1}{rgb}{0.529412,0.807843,1.000000}
  \definecolor{SkyBlue2}{rgb}{0.494118,0.752941,0.933333}
  \definecolor{SkyBlue3}{rgb}{0.423529,0.650980,0.803922}
  \definecolor{SkyBlue4}{rgb}{0.290196,0.439216,0.545098}
  \definecolor{LightSkyBlue1}{rgb}{0.690196,0.886275,1.000000}
  \definecolor{LightSkyBlue2}{rgb}{0.643137,0.827451,0.933333}
  \definecolor{LightSkyBlue3}{rgb}{0.552941,0.713726,0.803922}
  \definecolor{LightSkyBlue4}{rgb}{0.376471,0.482353,0.545098}
  \definecolor{SlateGray1}{rgb}{0.776471,0.886275,1.000000}
  \definecolor{SlateGray2}{rgb}{0.725490,0.827451,0.933333}
  \definecolor{SlateGray3}{rgb}{0.623529,0.713726,0.803922}
  \definecolor{SlateGray4}{rgb}{0.423529,0.482353,0.545098}
  \definecolor{LightSteelBlue1}{rgb}{0.792157,0.882353,1.000000}
  \definecolor{LightSteelBlue2}{rgb}{0.737255,0.823529,0.933333}
  \definecolor{LightSteelBlue3}{rgb}{0.635294,0.709804,0.803922}
  \definecolor{LightSteelBlue4}{rgb}{0.431373,0.482353,0.545098}
  \definecolor{LightBlue1}{rgb}{0.749020,0.937255,1.000000}
  \definecolor{LightBlue2}{rgb}{0.698039,0.874510,0.933333}
  \definecolor{LightBlue3}{rgb}{0.603922,0.752941,0.803922}
  \definecolor{LightBlue4}{rgb}{0.407843,0.513726,0.545098}
  \definecolor{LightCyan1}{rgb}{0.878431,1.000000,1.000000}
  \definecolor{LightCyan2}{rgb}{0.819608,0.933333,0.933333}
  \definecolor{LightCyan3}{rgb}{0.705882,0.803922,0.803922}
  \definecolor{LightCyan4}{rgb}{0.478431,0.545098,0.545098}
  \definecolor{PaleTurquoise1}{rgb}{0.733333,1.000000,1.000000}
  \definecolor{PaleTurquoise2}{rgb}{0.682353,0.933333,0.933333}
  \definecolor{PaleTurquoise3}{rgb}{0.588235,0.803922,0.803922}
  \definecolor{PaleTurquoise4}{rgb}{0.400000,0.545098,0.545098}
  \definecolor{CadetBlue1}{rgb}{0.596078,0.960784,1.000000}
  \definecolor{CadetBlue2}{rgb}{0.556863,0.898039,0.933333}
  \definecolor{CadetBlue3}{rgb}{0.478431,0.772549,0.803922}
  \definecolor{CadetBlue4}{rgb}{0.325490,0.525490,0.545098}
  \definecolor{turquoise1}{rgb}{0.000000,0.960784,1.000000}
  \definecolor{turquoise2}{rgb}{0.000000,0.898039,0.933333}
  \definecolor{turquoise3}{rgb}{0.000000,0.772549,0.803922}
  \definecolor{turquoise4}{rgb}{0.000000,0.525490,0.545098}
  \definecolor{cyan1}{rgb}{0.000000,1.000000,1.000000}
  \definecolor{cyan2}{rgb}{0.000000,0.933333,0.933333}
  \definecolor{cyan3}{rgb}{0.000000,0.803922,0.803922}
  \definecolor{cyan4}{rgb}{0.000000,0.545098,0.545098}
  \definecolor{DarkSlateGray1}{rgb}{0.592157,1.000000,1.000000}
  \definecolor{DarkSlateGray2}{rgb}{0.552941,0.933333,0.933333}
  \definecolor{DarkSlateGray3}{rgb}{0.474510,0.803922,0.803922}
  \definecolor{DarkSlateGray4}{rgb}{0.321569,0.545098,0.545098}
  \definecolor{aquamarine1}{rgb}{0.498039,1.000000,0.831373}
  \definecolor{aquamarine2}{rgb}{0.462745,0.933333,0.776471}
  \definecolor{aquamarine3}{rgb}{0.400000,0.803922,0.666667}
  \definecolor{aquamarine4}{rgb}{0.270588,0.545098,0.454902}
  \definecolor{DarkSeaGreen1}{rgb}{0.756863,1.000000,0.756863}
  \definecolor{DarkSeaGreen2}{rgb}{0.705882,0.933333,0.705882}
  \definecolor{DarkSeaGreen3}{rgb}{0.607843,0.803922,0.607843}
  \definecolor{DarkSeaGreen4}{rgb}{0.411765,0.545098,0.411765}
  \definecolor{SeaGreen1}{rgb}{0.329412,1.000000,0.623529}
  \definecolor{SeaGreen2}{rgb}{0.305882,0.933333,0.580392}
  \definecolor{SeaGreen3}{rgb}{0.262745,0.803922,0.501961}
  \definecolor{SeaGreen4}{rgb}{0.180392,0.545098,0.341176}
  \definecolor{PaleGreen1}{rgb}{0.603922,1.000000,0.603922}
  \definecolor{PaleGreen2}{rgb}{0.564706,0.933333,0.564706}
  \definecolor{PaleGreen3}{rgb}{0.486275,0.803922,0.486275}
  \definecolor{PaleGreen4}{rgb}{0.329412,0.545098,0.329412}
  \definecolor{SpringGreen1}{rgb}{0.000000,1.000000,0.498039}
  \definecolor{SpringGreen2}{rgb}{0.000000,0.933333,0.462745}
  \definecolor{SpringGreen3}{rgb}{0.000000,0.803922,0.400000}
  \definecolor{SpringGreen4}{rgb}{0.000000,0.545098,0.270588}
  \definecolor{green1}{rgb}{0.000000,1.000000,0.000000}
  \definecolor{green2}{rgb}{0.000000,0.933333,0.000000}
  \definecolor{green3}{rgb}{0.000000,0.803922,0.000000}
  \definecolor{green4}{rgb}{0.000000,0.545098,0.000000}
  \definecolor{chartreuse1}{rgb}{0.498039,1.000000,0.000000}
  \definecolor{chartreuse2}{rgb}{0.462745,0.933333,0.000000}
  \definecolor{chartreuse3}{rgb}{0.400000,0.803922,0.000000}
  \definecolor{chartreuse4}{rgb}{0.270588,0.545098,0.000000}
  \definecolor{OliveDrab1}{rgb}{0.752941,1.000000,0.243137}
  \definecolor{OliveDrab2}{rgb}{0.701961,0.933333,0.227451}
  \definecolor{OliveDrab3}{rgb}{0.603922,0.803922,0.196078}
  \definecolor{OliveDrab4}{rgb}{0.411765,0.545098,0.133333}
  \definecolor{DarkOliveGreen1}{rgb}{0.792157,1.000000,0.439216}
  \definecolor{DarkOliveGreen2}{rgb}{0.737255,0.933333,0.407843}
  \definecolor{DarkOliveGreen3}{rgb}{0.635294,0.803922,0.352941}
  \definecolor{DarkOliveGreen4}{rgb}{0.431373,0.545098,0.239216}
  \definecolor{khaki1}{rgb}{1.000000,0.964706,0.560784}
  \definecolor{khaki2}{rgb}{0.933333,0.901961,0.521569}
  \definecolor{khaki3}{rgb}{0.803922,0.776471,0.450980}
  \definecolor{khaki4}{rgb}{0.545098,0.525490,0.305882}
  \definecolor{LightGoldenrod1}{rgb}{1.000000,0.925490,0.545098}
  \definecolor{LightGoldenrod2}{rgb}{0.933333,0.862745,0.509804}
  \definecolor{LightGoldenrod3}{rgb}{0.803922,0.745098,0.439216}
  \definecolor{LightGoldenrod4}{rgb}{0.545098,0.505882,0.298039}
  \definecolor{LightYellow1}{rgb}{1.000000,1.000000,0.878431}
  \definecolor{LightYellow2}{rgb}{0.933333,0.933333,0.819608}
  \definecolor{LightYellow3}{rgb}{0.803922,0.803922,0.705882}
  \definecolor{LightYellow4}{rgb}{0.545098,0.545098,0.478431}
  \definecolor{yellow1}{rgb}{1.000000,1.000000,0.000000}
  \definecolor{yellow2}{rgb}{0.933333,0.933333,0.000000}
  \definecolor{yellow3}{rgb}{0.803922,0.803922,0.000000}
  \definecolor{yellow4}{rgb}{0.545098,0.545098,0.000000}
  \definecolor{gold1}{rgb}{1.000000,0.843137,0.000000}
  \definecolor{gold2}{rgb}{0.933333,0.788235,0.000000}
  \definecolor{gold3}{rgb}{0.803922,0.678431,0.000000}
  \definecolor{gold4}{rgb}{0.545098,0.458824,0.000000}
  \definecolor{goldenrod1}{rgb}{1.000000,0.756863,0.145098}
  \definecolor{goldenrod2}{rgb}{0.933333,0.705882,0.133333}
  \definecolor{goldenrod3}{rgb}{0.803922,0.607843,0.113725}
  \definecolor{goldenrod4}{rgb}{0.545098,0.411765,0.078431}
  \definecolor{DarkGoldenrod1}{rgb}{1.000000,0.725490,0.058824}
  \definecolor{DarkGoldenrod2}{rgb}{0.933333,0.678431,0.054902}
  \definecolor{DarkGoldenrod3}{rgb}{0.803922,0.584314,0.047059}
  \definecolor{DarkGoldenrod4}{rgb}{0.545098,0.396078,0.031373}
  \definecolor{RosyBrown1}{rgb}{1.000000,0.756863,0.756863}
  \definecolor{RosyBrown2}{rgb}{0.933333,0.705882,0.705882}
  \definecolor{RosyBrown3}{rgb}{0.803922,0.607843,0.607843}
  \definecolor{RosyBrown4}{rgb}{0.545098,0.411765,0.411765}
  \definecolor{IndianRed1}{rgb}{1.000000,0.415686,0.415686}
  \definecolor{IndianRed2}{rgb}{0.933333,0.388235,0.388235}
  \definecolor{IndianRed3}{rgb}{0.803922,0.333333,0.333333}
  \definecolor{IndianRed4}{rgb}{0.545098,0.227451,0.227451}
  \definecolor{sienna1}{rgb}{1.000000,0.509804,0.278431}
  \definecolor{sienna2}{rgb}{0.933333,0.474510,0.258824}
  \definecolor{sienna3}{rgb}{0.803922,0.407843,0.223529}
  \definecolor{sienna4}{rgb}{0.545098,0.278431,0.149020}
  \definecolor{burlywood1}{rgb}{1.000000,0.827451,0.607843}
  \definecolor{burlywood2}{rgb}{0.933333,0.772549,0.568627}
  \definecolor{burlywood3}{rgb}{0.803922,0.666667,0.490196}
  \definecolor{burlywood4}{rgb}{0.545098,0.450980,0.333333}
  \definecolor{wheat1}{rgb}{1.000000,0.905882,0.729412}
  \definecolor{wheat2}{rgb}{0.933333,0.847059,0.682353}
  \definecolor{wheat3}{rgb}{0.803922,0.729412,0.588235}
  \definecolor{wheat4}{rgb}{0.545098,0.494118,0.400000}
  \definecolor{tan1}{rgb}{1.000000,0.647059,0.309804}
  \definecolor{tan2}{rgb}{0.933333,0.603922,0.286275}
  \definecolor{tan3}{rgb}{0.803922,0.521569,0.247059}
  \definecolor{tan4}{rgb}{0.545098,0.352941,0.168627}
  \definecolor{chocolate1}{rgb}{1.000000,0.498039,0.141176}
  \definecolor{chocolate2}{rgb}{0.933333,0.462745,0.129412}
  \definecolor{chocolate3}{rgb}{0.803922,0.400000,0.113725}
  \definecolor{chocolate4}{rgb}{0.545098,0.270588,0.074510}
  \definecolor{firebrick1}{rgb}{1.000000,0.188235,0.188235}
  \definecolor{firebrick2}{rgb}{0.933333,0.172549,0.172549}
  \definecolor{firebrick3}{rgb}{0.803922,0.149020,0.149020}
  \definecolor{firebrick4}{rgb}{0.545098,0.101961,0.101961}
  \definecolor{brown1}{rgb}{1.000000,0.250980,0.250980}
  \definecolor{brown2}{rgb}{0.933333,0.231373,0.231373}
  \definecolor{brown3}{rgb}{0.803922,0.200000,0.200000}
  \definecolor{brown4}{rgb}{0.545098,0.137255,0.137255}
  \definecolor{salmon1}{rgb}{1.000000,0.549020,0.411765}
  \definecolor{salmon2}{rgb}{0.933333,0.509804,0.384314}
  \definecolor{salmon3}{rgb}{0.803922,0.439216,0.329412}
  \definecolor{salmon4}{rgb}{0.545098,0.298039,0.223529}
  \definecolor{LightSalmon1}{rgb}{1.000000,0.627451,0.478431}
  \definecolor{LightSalmon2}{rgb}{0.933333,0.584314,0.447059}
  \definecolor{LightSalmon3}{rgb}{0.803922,0.505882,0.384314}
  \definecolor{LightSalmon4}{rgb}{0.545098,0.341176,0.258824}
  \definecolor{orange1}{rgb}{1.000000,0.647059,0.000000}
  \definecolor{orange2}{rgb}{0.933333,0.603922,0.000000}
  \definecolor{orange3}{rgb}{0.803922,0.521569,0.000000}
  \definecolor{orange4}{rgb}{0.545098,0.352941,0.000000}
  \definecolor{DarkOrange1}{rgb}{1.000000,0.498039,0.000000}
  \definecolor{DarkOrange2}{rgb}{0.933333,0.462745,0.000000}
  \definecolor{DarkOrange3}{rgb}{0.803922,0.400000,0.000000}
  \definecolor{DarkOrange4}{rgb}{0.545098,0.270588,0.000000}
  \definecolor{coral1}{rgb}{1.000000,0.447059,0.337255}
  \definecolor{coral2}{rgb}{0.933333,0.415686,0.313726}
  \definecolor{coral3}{rgb}{0.803922,0.356863,0.270588}
  \definecolor{coral4}{rgb}{0.545098,0.243137,0.184314}
  \definecolor{tomato1}{rgb}{1.000000,0.388235,0.278431}
  \definecolor{tomato2}{rgb}{0.933333,0.360784,0.258824}
  \definecolor{tomato3}{rgb}{0.803922,0.309804,0.223529}
  \definecolor{tomato4}{rgb}{0.545098,0.211765,0.149020}
  \definecolor{OrangeRed1}{rgb}{1.000000,0.270588,0.000000}
  \definecolor{OrangeRed2}{rgb}{0.933333,0.250980,0.000000}
  \definecolor{OrangeRed3}{rgb}{0.803922,0.215686,0.000000}
  \definecolor{OrangeRed4}{rgb}{0.545098,0.145098,0.000000}
  \definecolor{red1}{rgb}{1.000000,0.000000,0.000000}
  \definecolor{red2}{rgb}{0.933333,0.000000,0.000000}
  \definecolor{red3}{rgb}{0.803922,0.000000,0.000000}
  \definecolor{red4}{rgb}{0.545098,0.000000,0.000000}
  \definecolor{DeepPink1}{rgb}{1.000000,0.078431,0.576471}
  \definecolor{DeepPink2}{rgb}{0.933333,0.070588,0.537255}
  \definecolor{DeepPink3}{rgb}{0.803922,0.062745,0.462745}
  \definecolor{DeepPink4}{rgb}{0.545098,0.039216,0.313726}
  \definecolor{HotPink1}{rgb}{1.000000,0.431373,0.705882}
  \definecolor{HotPink2}{rgb}{0.933333,0.415686,0.654902}
  \definecolor{HotPink3}{rgb}{0.803922,0.376471,0.564706}
  \definecolor{HotPink4}{rgb}{0.545098,0.227451,0.384314}
  \definecolor{pink1}{rgb}{1.000000,0.709804,0.772549}
  \definecolor{pink2}{rgb}{0.933333,0.662745,0.721569}
  \definecolor{pink3}{rgb}{0.803922,0.568627,0.619608}
  \definecolor{pink4}{rgb}{0.545098,0.388235,0.423529}
  \definecolor{LightPink1}{rgb}{1.000000,0.682353,0.725490}
  \definecolor{LightPink2}{rgb}{0.933333,0.635294,0.678431}
  \definecolor{LightPink3}{rgb}{0.803922,0.549020,0.584314}
  \definecolor{LightPink4}{rgb}{0.545098,0.372549,0.396078}
  \definecolor{PaleVioletRed1}{rgb}{1.000000,0.509804,0.670588}
  \definecolor{PaleVioletRed2}{rgb}{0.933333,0.474510,0.623529}
  \definecolor{PaleVioletRed3}{rgb}{0.803922,0.407843,0.537255}
  \definecolor{PaleVioletRed4}{rgb}{0.545098,0.278431,0.364706}
  \definecolor{maroon1}{rgb}{1.000000,0.203922,0.701961}
  \definecolor{maroon2}{rgb}{0.933333,0.188235,0.654902}
  \definecolor{maroon3}{rgb}{0.803922,0.160784,0.564706}
  \definecolor{maroon4}{rgb}{0.545098,0.109804,0.384314}
  \definecolor{VioletRed1}{rgb}{1.000000,0.243137,0.588235}
  \definecolor{VioletRed2}{rgb}{0.933333,0.227451,0.549020}
  \definecolor{VioletRed3}{rgb}{0.803922,0.196078,0.470588}
  \definecolor{VioletRed4}{rgb}{0.545098,0.133333,0.321569}
  \definecolor{magenta1}{rgb}{1.000000,0.000000,1.000000}
  \definecolor{magenta2}{rgb}{0.933333,0.000000,0.933333}
  \definecolor{magenta3}{rgb}{0.803922,0.000000,0.803922}
  \definecolor{magenta4}{rgb}{0.545098,0.000000,0.545098}
  \definecolor{orchid1}{rgb}{1.000000,0.513726,0.980392}
  \definecolor{orchid2}{rgb}{0.933333,0.478431,0.913725}
  \definecolor{orchid3}{rgb}{0.803922,0.411765,0.788235}
  \definecolor{orchid4}{rgb}{0.545098,0.278431,0.537255}
  \definecolor{plum1}{rgb}{1.000000,0.733333,1.000000}
  \definecolor{plum2}{rgb}{0.933333,0.682353,0.933333}
  \definecolor{plum3}{rgb}{0.803922,0.588235,0.803922}
  \definecolor{plum4}{rgb}{0.545098,0.400000,0.545098}
  \definecolor{MediumOrchid1}{rgb}{0.878431,0.400000,1.000000}
  \definecolor{MediumOrchid2}{rgb}{0.819608,0.372549,0.933333}
  \definecolor{MediumOrchid3}{rgb}{0.705882,0.321569,0.803922}
  \definecolor{MediumOrchid4}{rgb}{0.478431,0.215686,0.545098}
  \definecolor{DarkOrchid1}{rgb}{0.749020,0.243137,1.000000}
  \definecolor{DarkOrchid2}{rgb}{0.698039,0.227451,0.933333}
  \definecolor{DarkOrchid3}{rgb}{0.603922,0.196078,0.803922}
  \definecolor{DarkOrchid4}{rgb}{0.407843,0.133333,0.545098}
  \definecolor{purple1}{rgb}{0.607843,0.188235,1.000000}
  \definecolor{purple2}{rgb}{0.568627,0.172549,0.933333}
  \definecolor{purple3}{rgb}{0.490196,0.149020,0.803922}
  \definecolor{purple4}{rgb}{0.333333,0.101961,0.545098}
  \definecolor{MediumPurple1}{rgb}{0.670588,0.509804,1.000000}
  \definecolor{MediumPurple2}{rgb}{0.623529,0.474510,0.933333}
  \definecolor{MediumPurple3}{rgb}{0.537255,0.407843,0.803922}
  \definecolor{MediumPurple4}{rgb}{0.364706,0.278431,0.545098}
  \definecolor{thistle1}{rgb}{1.000000,0.882353,1.000000}
  \definecolor{thistle2}{rgb}{0.933333,0.823529,0.933333}
  \definecolor{thistle3}{rgb}{0.803922,0.709804,0.803922}
  \definecolor{thistle4}{rgb}{0.545098,0.482353,0.545098}
  \definecolor{gray0}{rgb}{0.000000,0.000000,0.000000}
  \definecolor{grey0}{rgb}{0.000000,0.000000,0.000000}
  \definecolor{gray1}{rgb}{0.011765,0.011765,0.011765}
  \definecolor{grey1}{rgb}{0.011765,0.011765,0.011765}
  \definecolor{gray2}{rgb}{0.019608,0.019608,0.019608}
  \definecolor{grey2}{rgb}{0.019608,0.019608,0.019608}
  \definecolor{gray3}{rgb}{0.031373,0.031373,0.031373}
  \definecolor{grey3}{rgb}{0.031373,0.031373,0.031373}
  \definecolor{gray4}{rgb}{0.039216,0.039216,0.039216}
  \definecolor{grey4}{rgb}{0.039216,0.039216,0.039216}
  \definecolor{gray5}{rgb}{0.050980,0.050980,0.050980}
  \definecolor{grey5}{rgb}{0.050980,0.050980,0.050980}
  \definecolor{gray6}{rgb}{0.058824,0.058824,0.058824}
  \definecolor{grey6}{rgb}{0.058824,0.058824,0.058824}
  \definecolor{gray7}{rgb}{0.070588,0.070588,0.070588}
  \definecolor{grey7}{rgb}{0.070588,0.070588,0.070588}
  \definecolor{gray8}{rgb}{0.078431,0.078431,0.078431}
  \definecolor{grey8}{rgb}{0.078431,0.078431,0.078431}
  \definecolor{gray9}{rgb}{0.090196,0.090196,0.090196}
  \definecolor{grey9}{rgb}{0.090196,0.090196,0.090196}
  \definecolor{gray10}{rgb}{0.101961,0.101961,0.101961}
  \definecolor{grey10}{rgb}{0.101961,0.101961,0.101961}
  \definecolor{gray11}{rgb}{0.109804,0.109804,0.109804}
  \definecolor{grey11}{rgb}{0.109804,0.109804,0.109804}
  \definecolor{gray12}{rgb}{0.121569,0.121569,0.121569}
  \definecolor{grey12}{rgb}{0.121569,0.121569,0.121569}
  \definecolor{gray13}{rgb}{0.129412,0.129412,0.129412}
  \definecolor{grey13}{rgb}{0.129412,0.129412,0.129412}
  \definecolor{gray14}{rgb}{0.141176,0.141176,0.141176}
  \definecolor{grey14}{rgb}{0.141176,0.141176,0.141176}
  \definecolor{gray15}{rgb}{0.149020,0.149020,0.149020}
  \definecolor{grey15}{rgb}{0.149020,0.149020,0.149020}
  \definecolor{gray16}{rgb}{0.160784,0.160784,0.160784}
  \definecolor{grey16}{rgb}{0.160784,0.160784,0.160784}
  \definecolor{gray17}{rgb}{0.168627,0.168627,0.168627}
  \definecolor{grey17}{rgb}{0.168627,0.168627,0.168627}
  \definecolor{gray18}{rgb}{0.180392,0.180392,0.180392}
  \definecolor{grey18}{rgb}{0.180392,0.180392,0.180392}
  \definecolor{gray19}{rgb}{0.188235,0.188235,0.188235}
  \definecolor{grey19}{rgb}{0.188235,0.188235,0.188235}
  \definecolor{gray20}{rgb}{0.200000,0.200000,0.200000}
  \definecolor{grey20}{rgb}{0.200000,0.200000,0.200000}
  \definecolor{gray21}{rgb}{0.211765,0.211765,0.211765}
  \definecolor{grey21}{rgb}{0.211765,0.211765,0.211765}
  \definecolor{gray22}{rgb}{0.219608,0.219608,0.219608}
  \definecolor{grey22}{rgb}{0.219608,0.219608,0.219608}
  \definecolor{gray23}{rgb}{0.231373,0.231373,0.231373}
  \definecolor{grey23}{rgb}{0.231373,0.231373,0.231373}
  \definecolor{gray24}{rgb}{0.239216,0.239216,0.239216}
  \definecolor{grey24}{rgb}{0.239216,0.239216,0.239216}
  \definecolor{gray25}{rgb}{0.250980,0.250980,0.250980}
  \definecolor{grey25}{rgb}{0.250980,0.250980,0.250980}
  \definecolor{gray26}{rgb}{0.258824,0.258824,0.258824}
  \definecolor{grey26}{rgb}{0.258824,0.258824,0.258824}
  \definecolor{gray27}{rgb}{0.270588,0.270588,0.270588}
  \definecolor{grey27}{rgb}{0.270588,0.270588,0.270588}
  \definecolor{gray28}{rgb}{0.278431,0.278431,0.278431}
  \definecolor{grey28}{rgb}{0.278431,0.278431,0.278431}
  \definecolor{gray29}{rgb}{0.290196,0.290196,0.290196}
  \definecolor{grey29}{rgb}{0.290196,0.290196,0.290196}
  \definecolor{gray30}{rgb}{0.301961,0.301961,0.301961}
  \definecolor{grey30}{rgb}{0.301961,0.301961,0.301961}
  \definecolor{gray31}{rgb}{0.309804,0.309804,0.309804}
  \definecolor{grey31}{rgb}{0.309804,0.309804,0.309804}
  \definecolor{gray32}{rgb}{0.321569,0.321569,0.321569}
  \definecolor{grey32}{rgb}{0.321569,0.321569,0.321569}
  \definecolor{gray33}{rgb}{0.329412,0.329412,0.329412}
  \definecolor{grey33}{rgb}{0.329412,0.329412,0.329412}
  \definecolor{gray34}{rgb}{0.341176,0.341176,0.341176}
  \definecolor{grey34}{rgb}{0.341176,0.341176,0.341176}
  \definecolor{gray35}{rgb}{0.349020,0.349020,0.349020}
  \definecolor{grey35}{rgb}{0.349020,0.349020,0.349020}
  \definecolor{gray36}{rgb}{0.360784,0.360784,0.360784}
  \definecolor{grey36}{rgb}{0.360784,0.360784,0.360784}
  \definecolor{gray37}{rgb}{0.368627,0.368627,0.368627}
  \definecolor{grey37}{rgb}{0.368627,0.368627,0.368627}
  \definecolor{gray38}{rgb}{0.380392,0.380392,0.380392}
  \definecolor{grey38}{rgb}{0.380392,0.380392,0.380392}
  \definecolor{gray39}{rgb}{0.388235,0.388235,0.388235}
  \definecolor{grey39}{rgb}{0.388235,0.388235,0.388235}
  \definecolor{gray40}{rgb}{0.400000,0.400000,0.400000}
  \definecolor{grey40}{rgb}{0.400000,0.400000,0.400000}
  \definecolor{gray41}{rgb}{0.411765,0.411765,0.411765}
  \definecolor{grey41}{rgb}{0.411765,0.411765,0.411765}
  \definecolor{gray42}{rgb}{0.419608,0.419608,0.419608}
  \definecolor{grey42}{rgb}{0.419608,0.419608,0.419608}
  \definecolor{gray43}{rgb}{0.431373,0.431373,0.431373}
  \definecolor{grey43}{rgb}{0.431373,0.431373,0.431373}
  \definecolor{gray44}{rgb}{0.439216,0.439216,0.439216}
  \definecolor{grey44}{rgb}{0.439216,0.439216,0.439216}
  \definecolor{gray45}{rgb}{0.450980,0.450980,0.450980}
  \definecolor{grey45}{rgb}{0.450980,0.450980,0.450980}
  \definecolor{gray46}{rgb}{0.458824,0.458824,0.458824}
  \definecolor{grey46}{rgb}{0.458824,0.458824,0.458824}
  \definecolor{gray47}{rgb}{0.470588,0.470588,0.470588}
  \definecolor{grey47}{rgb}{0.470588,0.470588,0.470588}
  \definecolor{gray48}{rgb}{0.478431,0.478431,0.478431}
  \definecolor{grey48}{rgb}{0.478431,0.478431,0.478431}
  \definecolor{gray49}{rgb}{0.490196,0.490196,0.490196}
  \definecolor{grey49}{rgb}{0.490196,0.490196,0.490196}
  \definecolor{gray50}{rgb}{0.498039,0.498039,0.498039}
  \definecolor{grey50}{rgb}{0.498039,0.498039,0.498039}
  \definecolor{gray51}{rgb}{0.509804,0.509804,0.509804}
  \definecolor{grey51}{rgb}{0.509804,0.509804,0.509804}
  \definecolor{gray52}{rgb}{0.521569,0.521569,0.521569}
  \definecolor{grey52}{rgb}{0.521569,0.521569,0.521569}
  \definecolor{gray53}{rgb}{0.529412,0.529412,0.529412}
  \definecolor{grey53}{rgb}{0.529412,0.529412,0.529412}
  \definecolor{gray54}{rgb}{0.541176,0.541176,0.541176}
  \definecolor{grey54}{rgb}{0.541176,0.541176,0.541176}
  \definecolor{gray55}{rgb}{0.549020,0.549020,0.549020}
  \definecolor{grey55}{rgb}{0.549020,0.549020,0.549020}
  \definecolor{gray56}{rgb}{0.560784,0.560784,0.560784}
  \definecolor{grey56}{rgb}{0.560784,0.560784,0.560784}
  \definecolor{gray57}{rgb}{0.568627,0.568627,0.568627}
  \definecolor{grey57}{rgb}{0.568627,0.568627,0.568627}
  \definecolor{gray58}{rgb}{0.580392,0.580392,0.580392}
  \definecolor{grey58}{rgb}{0.580392,0.580392,0.580392}
  \definecolor{gray59}{rgb}{0.588235,0.588235,0.588235}
  \definecolor{grey59}{rgb}{0.588235,0.588235,0.588235}
  \definecolor{gray60}{rgb}{0.600000,0.600000,0.600000}
  \definecolor{grey60}{rgb}{0.600000,0.600000,0.600000}
  \definecolor{gray61}{rgb}{0.611765,0.611765,0.611765}
  \definecolor{grey61}{rgb}{0.611765,0.611765,0.611765}
  \definecolor{gray62}{rgb}{0.619608,0.619608,0.619608}
  \definecolor{grey62}{rgb}{0.619608,0.619608,0.619608}
  \definecolor{gray63}{rgb}{0.631373,0.631373,0.631373}
  \definecolor{grey63}{rgb}{0.631373,0.631373,0.631373}
  \definecolor{gray64}{rgb}{0.639216,0.639216,0.639216}
  \definecolor{grey64}{rgb}{0.639216,0.639216,0.639216}
  \definecolor{gray65}{rgb}{0.650980,0.650980,0.650980}
  \definecolor{grey65}{rgb}{0.650980,0.650980,0.650980}
  \definecolor{gray66}{rgb}{0.658824,0.658824,0.658824}
  \definecolor{grey66}{rgb}{0.658824,0.658824,0.658824}
  \definecolor{gray67}{rgb}{0.670588,0.670588,0.670588}
  \definecolor{grey67}{rgb}{0.670588,0.670588,0.670588}
  \definecolor{gray68}{rgb}{0.678431,0.678431,0.678431}
  \definecolor{grey68}{rgb}{0.678431,0.678431,0.678431}
  \definecolor{gray69}{rgb}{0.690196,0.690196,0.690196}
  \definecolor{grey69}{rgb}{0.690196,0.690196,0.690196}
  \definecolor{gray70}{rgb}{0.701961,0.701961,0.701961}
  \definecolor{grey70}{rgb}{0.701961,0.701961,0.701961}
  \definecolor{gray71}{rgb}{0.709804,0.709804,0.709804}
  \definecolor{grey71}{rgb}{0.709804,0.709804,0.709804}
  \definecolor{gray72}{rgb}{0.721569,0.721569,0.721569}
  \definecolor{grey72}{rgb}{0.721569,0.721569,0.721569}
  \definecolor{gray73}{rgb}{0.729412,0.729412,0.729412}
  \definecolor{grey73}{rgb}{0.729412,0.729412,0.729412}
  \definecolor{gray74}{rgb}{0.741176,0.741176,0.741176}
  \definecolor{grey74}{rgb}{0.741176,0.741176,0.741176}
  \definecolor{gray75}{rgb}{0.749020,0.749020,0.749020}
  \definecolor{grey75}{rgb}{0.749020,0.749020,0.749020}
  \definecolor{gray76}{rgb}{0.760784,0.760784,0.760784}
  \definecolor{grey76}{rgb}{0.760784,0.760784,0.760784}
  \definecolor{gray77}{rgb}{0.768627,0.768627,0.768627}
  \definecolor{grey77}{rgb}{0.768627,0.768627,0.768627}
  \definecolor{gray78}{rgb}{0.780392,0.780392,0.780392}
  \definecolor{grey78}{rgb}{0.780392,0.780392,0.780392}
  \definecolor{gray79}{rgb}{0.788235,0.788235,0.788235}
  \definecolor{grey79}{rgb}{0.788235,0.788235,0.788235}
  \definecolor{gray80}{rgb}{0.800000,0.800000,0.800000}
  \definecolor{grey80}{rgb}{0.800000,0.800000,0.800000}
  \definecolor{gray81}{rgb}{0.811765,0.811765,0.811765}
  \definecolor{grey81}{rgb}{0.811765,0.811765,0.811765}
  \definecolor{gray82}{rgb}{0.819608,0.819608,0.819608}
  \definecolor{grey82}{rgb}{0.819608,0.819608,0.819608}
  \definecolor{gray83}{rgb}{0.831373,0.831373,0.831373}
  \definecolor{grey83}{rgb}{0.831373,0.831373,0.831373}
  \definecolor{gray84}{rgb}{0.839216,0.839216,0.839216}
  \definecolor{grey84}{rgb}{0.839216,0.839216,0.839216}
  \definecolor{gray85}{rgb}{0.850980,0.850980,0.850980}
  \definecolor{grey85}{rgb}{0.850980,0.850980,0.850980}
  \definecolor{gray86}{rgb}{0.858824,0.858824,0.858824}
  \definecolor{grey86}{rgb}{0.858824,0.858824,0.858824}
  \definecolor{gray87}{rgb}{0.870588,0.870588,0.870588}
  \definecolor{grey87}{rgb}{0.870588,0.870588,0.870588}
  \definecolor{gray88}{rgb}{0.878431,0.878431,0.878431}
  \definecolor{grey88}{rgb}{0.878431,0.878431,0.878431}
  \definecolor{gray89}{rgb}{0.890196,0.890196,0.890196}
  \definecolor{grey89}{rgb}{0.890196,0.890196,0.890196}
  \definecolor{gray90}{rgb}{0.898039,0.898039,0.898039}
  \definecolor{grey90}{rgb}{0.898039,0.898039,0.898039}
  \definecolor{gray91}{rgb}{0.909804,0.909804,0.909804}
  \definecolor{grey91}{rgb}{0.909804,0.909804,0.909804}
  \definecolor{gray92}{rgb}{0.921569,0.921569,0.921569}
  \definecolor{grey92}{rgb}{0.921569,0.921569,0.921569}
  \definecolor{gray93}{rgb}{0.929412,0.929412,0.929412}
  \definecolor{grey93}{rgb}{0.929412,0.929412,0.929412}
  \definecolor{gray94}{rgb}{0.941176,0.941176,0.941176}
  \definecolor{grey94}{rgb}{0.941176,0.941176,0.941176}
  \definecolor{gray95}{rgb}{0.949020,0.949020,0.949020}
  \definecolor{grey95}{rgb}{0.949020,0.949020,0.949020}
  \definecolor{gray96}{rgb}{0.960784,0.960784,0.960784}
  \definecolor{grey96}{rgb}{0.960784,0.960784,0.960784}
  \definecolor{gray97}{rgb}{0.968627,0.968627,0.968627}
  \definecolor{grey97}{rgb}{0.968627,0.968627,0.968627}
  \definecolor{gray98}{rgb}{0.980392,0.980392,0.980392}
  \definecolor{grey98}{rgb}{0.980392,0.980392,0.980392}
  \definecolor{gray99}{rgb}{0.988235,0.988235,0.988235}
  \definecolor{grey99}{rgb}{0.988235,0.988235,0.988235}
  \definecolor{gray100}{rgb}{1.000000,1.000000,1.000000}
  \definecolor{grey100}{rgb}{1.000000,1.000000,1.000000}
  \definecolor{dark grey}{rgb}{0.662745,0.662745,0.662745}
  \definecolor{DarkGrey}{rgb}{0.662745,0.662745,0.662745}
  \definecolor{dark gray}{rgb}{0.662745,0.662745,0.662745}
  \definecolor{DarkGray}{rgb}{0.662745,0.662745,0.662745}
  \definecolor{dark blue}{rgb}{0.000000,0.000000,0.545098}
  \definecolor{DarkBlue}{rgb}{0.000000,0.000000,0.545098}
  \definecolor{dark cyan}{rgb}{0.000000,0.545098,0.545098}
  \definecolor{DarkCyan}{rgb}{0.000000,0.545098,0.545098}
  \definecolor{dark magenta}{rgb}{0.545098,0.000000,0.545098}
  \definecolor{DarkMagenta}{rgb}{0.545098,0.000000,0.545098}
  \definecolor{dark red}{rgb}{0.545098,0.000000,0.000000}
  \definecolor{DarkRed}{rgb}{0.545098,0.000000,0.000000}
  \definecolor{light green}{rgb}{0.564706,0.933333,0.564706}
  \definecolor{LightGreen}{rgb}{0.564706,0.933333,0.564706}

\newcommand{\kms}{km~s$^{-1}$}
\newcommand{\Msun}{M$_{\sun}$}

\title[Small galaxies: WDM vs CDM]{Faint dwarfs as a test of DM models: WDM vs. CDM}
\author[F. Governato et al.]{F.Governato$^{1}$,\thanks{E-mail:(FG); fabiog@astro.washington.edu} 
        D.Weisz$^{1,2,3}$,
        A.Pontzen$^{4}$,
        S.Loebman$^{5}$,
        D.Reed$^{6}$,
        A.\,M.Brooks$^{7}$,
 \newauthor
        P.Behroozi$^{8}$,
        C.Christensen$^{9}$,
        P.Madau$^{2}$,
        L.Mayer$^{10}$,
        S.Shen$^{2}$, 
       M.Walker$^{11}$,
\newauthor
        T.Quinn$^{1}$, B.W.Keller$^{12}$
        and J.Wadsley$^{12}$ \\
$^{1}$Astronomy Department, University of Washington, Box 351580, Seattle, WA, 98195-1580 \\
$^{2}$Dept. of Astronomy, University of California at Santa Cruz, 1156 High Street, Santa Cruz, CA, 95064 USA; drw@ucsc.edu\\
$^{3}$Hubble Fellow\\
$^{4}$ UCL, Department of Physics \& Astronomy, Gower Place, London WC1E 6BT, UK\\
$^{5}$ Postdoctoral Fellow, Michigan Society of Fellows, Univ. of Michigan, Astronomy Dept., 830 Dennison 500 Church St. Ann Arbor, MI 48109 \\
$^{6}$ Institut de Ci\`encies de l'Espai (ICE, IEEC-CSIC), 08193 Bellaterra (Barcelona), Spain \\
$^{7}$ Dept. of Physics \& Astronomy Rutgers Univ. 136 Frelinghuysen Rd, Piscataway, NJ 08854  \\
$^{8}$Giacconi Postdoctoral Fellow, Space Telescope Institute, Baltimore, MD\\
$^{9}$Department of Astronomy, University of Arizona, 933 North Cherry Avenue, Rm. N204, Tucson, AZ 85721-0065, USA\\
$^{10}$ Center for Theoretical Astrophysics and Cosmology, Institute for Computational Science, Zurich, Switzerland\\
$^{11}$ McWilliams Center for Cosmology, Carnegie Mellon University, Pittsburgh, PA 15213 \\
$^{12}$Dept. of Physics and Astronomy, McMaster Univ., Hamilton, Ontario, L88 4M1, CA\\
}

\begin{document}

\date{Submitted  March 1st,2014}

\pagerange{\pageref{firstpage}--\pageref{lastpage}} \pubyear{2002}

\maketitle

\label{firstpage}
\voffset=-0.9cm   
\begin{abstract}

  We use high resolution Hydro$+$N-Body cosmological simulations to
  compare the assembly and evolution of a small field dwarf (stellar
  mass $\sim$ 10$^{6-7}$\Msun, total mass 10$^{10}$\Msun) in $\Lambda$
  dominated CDM and 2keV WDM cosmologies. We find that star formation
  (SF) in the WDM model is reduced and delayed by 1-2 Gyr relative to
  the CDM model, independently of the details of SF and feedback.
  Independent of the DM model, but proportionally to the SF
  efficiency, gas outflows lower the central mass density through
  `dynamical heating', such that all realizations have circular
  velocities $<$ 20 \kms at 500~pc, in agreement with local kinematic
  constraints. As a result of dynamical heating, older stars are less
  centrally concentrated than younger stars, similar to stellar
  population gradients observed in nearby dwarf galaxies. 
    Introducing an important diagnostic of SF and feedback models, we
  translate our simulations into artificial color-magnitude diagrams
  and star formation histories in order to directly compare to
  available observations.  The simulated galaxies formed most of their
  stars in many $\sim$10 Myr long bursts. The CDM galaxy has a global
  SFH, HI abundance and Fe/H and alpha-elements distribution well
  matched to current observations of dwarf galaxies.  These results
  highlight the importance of directly including `baryon physics' in
  simulations when 1) comparing predictions of galaxy formation models
  with the kinematics and number density of local dwarf galaxies and
  2) differentiating between CDM and non-standard models with
  different DM or power spectra.

 \end{abstract}

\begin{keywords}
Galaxies: formation -- Cosmology -- Hydrodynamics, dwarf galaxies, Galaxy Formation.
\end{keywords}

\section{Introduction}

Rapid advancements in numerical and semianalytic techniques and
modeling of star formation (SF) and feedback processes have lead to
studies of how dark matter and baryons assemble into cosmic structures
within the $\Lambda$ Cold Dark Matter (CDM) scenario
\citep{robertson04, dutton09fb,frenkwhite12,bensonwdm13}. At the same
time, strong observational evidence has shown how a large fraction of
the energy from stellar and quasi-stellar processes couple to the
surrounding interstellar medium at all redshifts
\citep{mcquinn10,vanderwel11}.  SF--related energy coupling (or
`feedback') to the interstellar medium (ISM) is necessary to regulate
SF and quench the formation of bulges
\citep{brook11,behroozi13b,hopkins13,munshi13}.

Recent studies have shown how feedback not only regulates star
formation, but directly affects the DM distribution. Rapid (faster
than the local dynamical time) removal of gas at sub-kpc scales causes
rapid and repeated fluctuations in the gravitational potential,
resulting in energy transfer to the dark matter (DM) component,
significantly lowering the central density of DM halos
\citep{mashchenko08,G10,G12,PG12,
  martizzi13,dicintio13,review14}.  The bursty
nature of SF that leads to lower galaxy central densities has strong
support from observations \citep{kauffmann14} and when properly
incorporated into simulations, results in galaxies with photometric
and kinematic properties close to those of real ones
\citep{G09,oh11,christensen14,kassin14}.  Combined estimates of the
abundance and internal mass distribution of faint field galaxies are
now constraining models of galaxy formation in CDM cosmologies, which
seem to over predict the number abundance and central density of halos
with mass $<$10$^{10}$M$\odot$
\citep{moore98,walker10,papastergis11,gk14,kirby14,klypin14}.
Motivated by discrepancies between CDM predictions and observational
evidence, theoretical models have also started to explore how
`non--standard' models would affect the assembly and inner structure
of galaxies.  Together with dynamical Dark Energy scenarios
\citep{penzo14}, recent works have explored physically motivated DM
models that specifically alter the number of small DM halos or lower
their central DM density, such as warm DM (WDM) and Self Interacting
DM (SIDM) \citep{spergel00,buckley10,zurek13,zavala13,vogelsberger14},

By damping density perturbations below a streaming length dependent on
the particle mass, WDM models reduce the abundance of small halos and
delay the collapse of halos above it \citep{calura14}. Introducing
WDM then lowers the slope of the faint end of the galaxy luminosity
function \citep{menci12} and the abundance of of Milky Way
satellites \citep{lovell12}. Analyses of the first simulations explicitly
including WDM and gas dynamics have also suggested that galaxy
formation in WDM could reduce the mass of stellar bulges
\citep{G04,maccioWDM}, by reducing the number of mergers and
interactions. A 2keV WDM scenario can also be used as a maximal case
for a class of models advocating for a non-trivial, 'rolling index' power
spectrum of density perturbations which reduces power at dwarf scales
\citep{gk14b,hazra14}, but without the abrupt filtering scale
associated with WDM.

\setlength{\textfloatsep}{0.5cm}

\begin{table}

\centering
\begin{tabular}{|p{0.2cm} p{2.7cm} p{1.7cm} p{1.6cm}|}
\hline
Run        & ~~~~SF Model & Stellar Mass  & HI mass  \\
 ID        &          &  CDM/WDM      & CDM/WDM                         \\
\hline
g1 & metal cooling, UV  & 4.8/1.3~$\times$10$^6$  & ~~17/36~$\times$10$^6$\Msun \\
g3 & g1$+$HI self--shielding & 8.0/3.0~$\times$10$^6$ & ~9.2/14~$\times$10$^6$\Msun\\
g5 & g3$+$Early Feedback &  3.1/0.6~$\times$10$^6$ & ~5.0/7.2~$\times$10$^6$\Msun\\

\hline
\end{tabular}
\caption{{\it The SF and feedback models  used in the simulations vs the z=0 stellar and HI masses (in \Msun)} for both CDM and WDM cosmologies.
  z$=0$ stellar masses were  measured within 2.5kpc. The WDM realizations  have significantly lower stellar masses and higher HI/stellar mass ratios,
  independently of the SF implementation.}
\end{table}

Unfortunately, most of the astrophysically driven support for
non-standard DM models has come from simulations that neglected the
complexities of `baryon physics' and followed only the the assembly of
the DM component. This approach has weak predicting power when it
comes to comparisons with the observed properties of real galaxies and
does not fully take advantage of available constraints.  
  Specifically, space-based surveys such as the ACS Nearby Galaxy
Survey Treasury \citep[or ANGST,][]{weisz11,ANGSTHI}, that map
the spatial distribution and colors of resolved stellar populations
allow us to reconstruct the detailed SFHs and metal enrichment
histories of the gas and stellar content of nearby dwarf galaxies
\citep{ANGSTSFR,weisz14}. Surveys as the Sloan Digital Sky Survey
(SDSS) provide information on the stellar content and instantaneous SF
of dwarf galaxies \citep{kauffmann14}. Data from these combined
observations provide new powerful tests for both SF and dark matter
models.

As a CDM model coupled to gas outflows has been shown to lower the
central densities of galaxies and quench SF in small systems, `baryon
physics' could then remove the need for non--standard DM models, which
were introduced to solve the same problems.  However, the very low SF
efficiency in galaxies forming less than 10$^7$\Msun in stars may not
provide sufficient energy to create DM cores in their host halos
\cite{G12,penarrubia12,dicintio13}.  The above
arguments strongly suggest that an explicit inclusion of baryon
physics in simulations is necessary to investigate the separate the
effects of SF and feedback from the possible unique signature of a
non-standard DM model, high resolution predictions of kinematics and
SFHs needs to be extended to the properties of the smallest field
dwarfs. In this work we will analyze very high resolution simulations
of dwarf galaxies in CDM vs WDM cosmologies, to compare their detailed
SFHs together with the evolution of the DM distribution at the centre
of their host halos.  In \S 2 and \S3 we describe the simulations and
the code used, In \S 4 we discuss the assembly of the baryon and DM
components in the two cosmologies.  In \S 5 we discuss the properties
of the stellar component, in \S 6 will describe the mass and DM
distribution at their center and summarize the results in \S 7. We
provide additional details on our analysis and artificial observations
in Appendix A.

\section{The Tree+SPH code ChaNGA: a new Version of Gasoline}

The simulations discussed in this paper were run in a full
cosmological context down to a redshift of zero using the $N$--body
Treecode $+$ Smoothed Particle Hydrodynamics (SPH) code {\sc ChaNGa}
\citep{changa08,changa13}.  main novelty consists in the dynamic load
balancing and computation/communication overlap provided by the
CHARM++ run-time system \citep{kale96}. {\sc{ChaNGA}} scales up to
100,000$+$ cores and down to only 5,000 particles per core. The
  communication and multistepping approach, together with extensive
  scaling tests and a general description of the code are presented in
  \citep{menon14}. {\sc ChaNGa} uses the same hydrodynamic and
physics modules previously introduced in {\sc{GASOLINE}}
\citep{wadsley04,stinson06,wadsley08}  together with some
  improvements to the SPH approach that have already described in
  the literature (see below).  The developers of {\sc{ChaNGA}}
  and {\sc{Gasoline}} are part of the {\sc{AGORA}} group, a research
  collaboration with the goal of comparing the implementation of
  hydrodynamics in cosmological codes \citep{AGORA}.

\begin{figure}
\includegraphics[width=85mm]{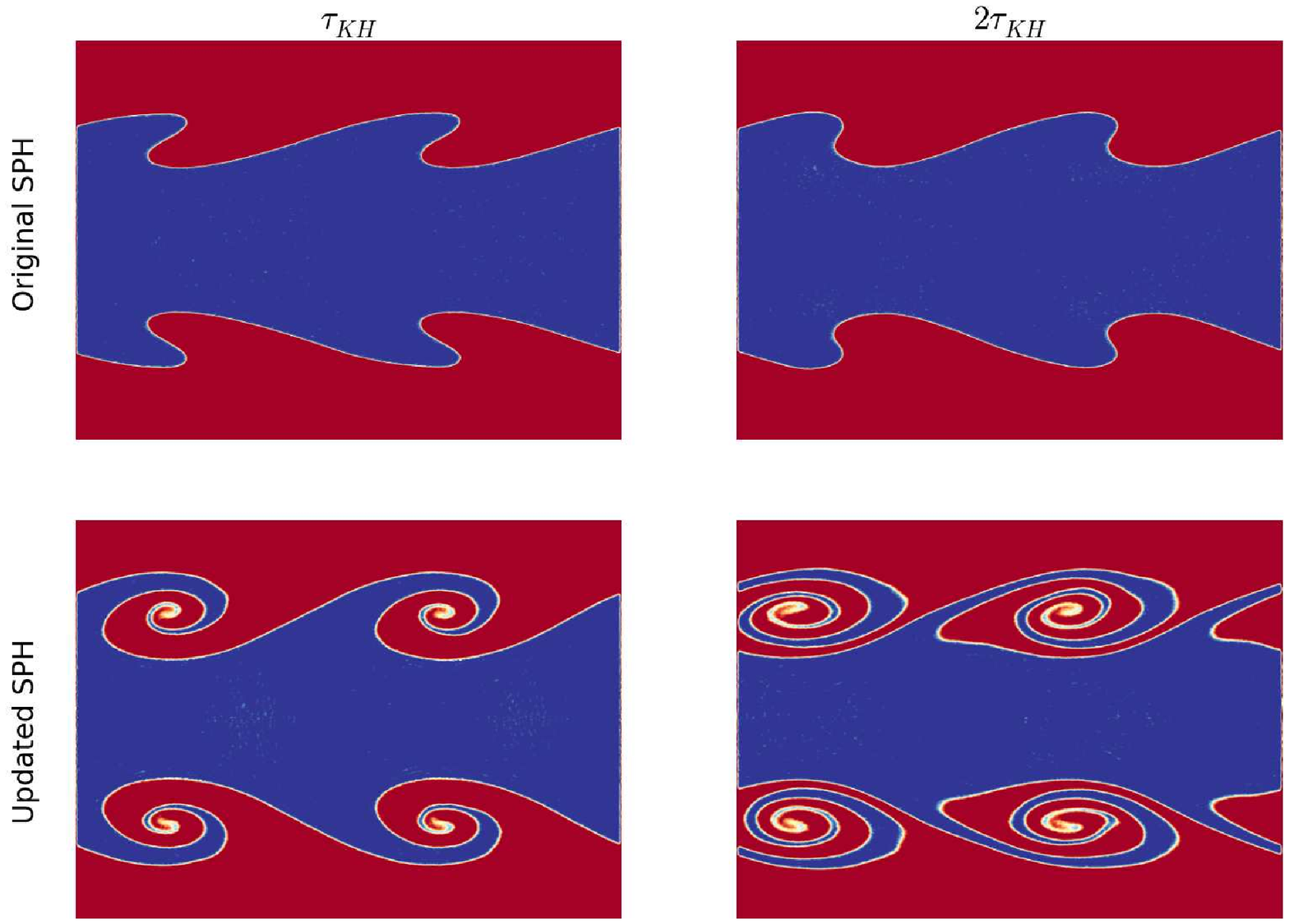}
\caption{{\it The growth of a Kelvin-Helmholtz (KH) instability
    at 1 and 2 KH times ($\tau_{KH}$)}. The top row shows how, using the
    "traditional" SPH method used in ChaNGa and GASOLINE
    \citep{wadsley04} spurious surface tension acts to suppress the
    growth of the KH instability.  With the inclusion of a turbulent
    diffusion term to properly address mixing \citep{shen10} and an
    alternate force expression \citep{ritchie01} this problem is
    alleviated, and instabilities can grow as they should.  This
    updated SPH method captures mixing in multi-phase gas (an
    important feature for galaxy simulations) much more accurately.
     }
\label{fig1}
\end{figure}

Recent works \citep{agertz10} have highlighted how the
modeling of pressure gradients in early SPH implementations
\citep{gadget2} created artificial surface tension between cold dense
gas and the hot, more dilute phase.  A first step to improve physics
at hot-cold interfaces was the addition in {\sc{ChaNGA/Gasoline}} of a
physically-motivated model for thermal diffusion, described in
\cite{shen10}.  We note that all codes, including SPH, have low
effective Reynolds numbers compared to real astrophysical flows and
thus the required turbulent mixing must often be treated
explicitly. We verified that results (and especially the metallicity
distribution of stellar populations) do not depend on the details of
the mixing model.  A key change in {\sc{ChaNGA/Gasoline}} is the
elimination of artificial gas surface tension through the use of a
geometric density mean in the SPH force expression:
$(P_i+P_j)/(\rho_i\rho_j)$ in place of $P_i/\rho_i^2+P_j/\rho_j^2$
where $P_i$ and $\rho_i$ are particle pressures and densities
respectively  \citep{ritchie01}. This update better simulates
shearing flows with Kelvin-Helmholtz instabilities (Fig~\ref{fig1}) and
 avoids the artificial formation of cold blobs in smooth flows
\citep{menon14}.  Detailed tests of the updated SPH, as implemented
in ChaNGa and Gasoline will be presented in Wadsley et al. (2014 in
prep.).
Equivalent SPH improvements have been proposed and shown to
alleviate artificial surface tension and correctly model strong shocks
\citep{readsph10,saitohsph,hopkinsSPH}.  Note that some authors
preferentially track entropy rather than energy but all these newer
implementations use a geometric density mean for the forces.  Further
independent testing of these approaches with successful results on a
wide range of problems has been presented by \cite{kawata13}. Thus,
there appears to be a growing consensus on the essential ingredients
for a modern SPH implementation that can correctly treat the transonic
and supersonic flows ubiquitous in astrophysics.

\section{CDM and WDM Simulations and Star Formation Parameters}
\label{sims}

We simulated a single halo in a filamentary region of average density
identified in a low resolution, 25 Mpc per side uniform volume
simulation that assumed a $\Lambda$CDM cosmology ($\Omega_0$ $=$0.24
$\Lambda$ $=$0.76, $\sigma_8$ $=$0.77 and n$=$0.96).  The halo was
then resimulated at high resolution in CDM and WDM cosmologies using
the ‘zoom-in‘ technique \citep{katz93}. The gravitational force spline
softening length was set at 60~pc \citep{power03} and a the minimum
gas smoothing length (roughly corresponding to the nominal resolution
for hydro forces) at 6pc. DM, gas and star particles (when first
created) have masses of 6650, 1410, and 422 \Msun, respectively. With
a time evolving mass due to stellar evolution \citep{wadsley04},
metallicity of the parent gas particle and fixed IMF, each star
particle is effectively a single stellar population.  The resolution
of these simulations is equivalent to a uniform grid of 4096$^3$
particles over the entire original volume, and is among the highest
published for cosmological simulations carried to z$=$0. By preserving
the wave phases in the two simulations, the 'zoomed in' approach
allows us to isolate the effect of changing the power spectrum on the
assembly of individual galaxies and preserving the large scale
structure (see also \cite{G04} and \cite{maccioWDM} for a similar
approach).

The WDM cosmology adopted here assumes a 2 keV particle and the
transfer function was obtained following the fitting formula in
\cite{viel05}. This is the 'warmest' model marginally allowed by
current constraints \citep{schneider14,viel13}; it has been
chosen to emphasize the possible differences between CDM and other
models with reduced small scales power. The thermal component of the
initial velocity field is negligible at the resolved scales. The
accuracy of the WDM halo mass function has been tested by comparing
with results from \cite{schneider12}.  Due to its lack of power at
smaller scales, the WDM simulation starts at z$=$119, while its CDM
counterpart starts at z$=$129.  Following \cite{lovell12}, we expect
numerical fragmentation effects due finite force and mass resolution
\citep{angulo13} to become negligible at halo masses below a
few times 10$^7$ \Msun. Empirically, we verified that the typical
  artificial halos originated by fragmentation have typically 300 DM
  particles within their virial radius, corresponding to $\sim 2
  \times$ 10$^5$ \Msun. These small halos contain no gas and oftentimes do not
  survive the tidal field of larger host halos.  We also verified that
  no stars form in halos with mass below 10$^{7.5}$ \Msun at any time
  in the simulation, providing strong support to the notion that
  spurious WDM halos do not affect our results.

\begin{figure*}
\includegraphics[width=160mm]{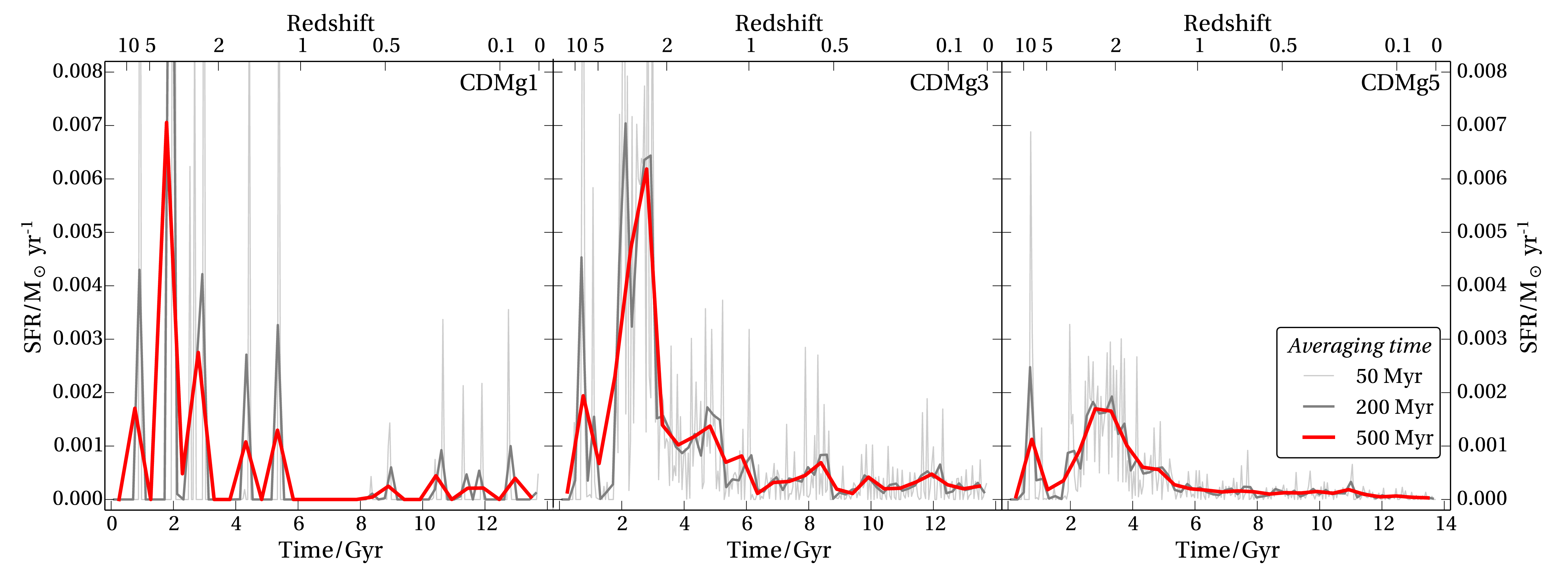}

\caption{ {\it The SFR of the galaxy in the CDM cosmologies with
    different time binsize.}  SFR is measured including all the stars
  within the central 1kpc at z$=$0.  The lack of self-shielding in
  CDMg1 gives sparse and very bursty SF. The SFH histories of the g3
  (which includes self-shielding) and g5 (self-shielding and `early'
  feedback) runs, are composed of many, short ($\sim$ 10 Myrs) bursty
  events. During these bursts, SF increases by a factor ranging from 1
  to 10 compared to the local time average. This behaviour is typical
  in the SFHs of dwarf galaxies obtained from SDSS
  \citep{kauffmann14}.  The SFH of runs g3 and g5 become smoother if
  averaged over longer time intervals that approach what is possible
  in the deepest observations of LG dwarf galaxies. In all runs the SF
  events create outflows that lower the central DM density of the
  host halo (Fig.~\ref{fig7} \& Fig.~\ref{fig8}).}
\label{fig2}
\end{figure*}

\begin{table}

\centering
\begin{tabular}{|p{1.9cm} p{1.7cm} p{1.cm} p{2.3cm}|}
\hline
DM    & DM/Gas & Gravtn. &  z$=$0 Halo Mass  \\
model &  particle mass &  Softening      &   (DM-only run)   \\
\hline
CDM(WDM) & 6650/1410   & ~~~60  & 1.16(1.12) $\times$10$^{10}$\\

\hline
\end{tabular}
\caption{{\it Simulation and Halos Properties.} 
  Column (2)  lists the mass (in \Msun)  of individual dark matter and star particles in the high resolution region.   Column (3) shows  $\epsilon$, the spline gravitational force softening  (4) lists the halo mass at z$=$0 for the DM-only runs (in \Msun). The minimum smoothing length for the SPH calculation is set to 0.1 the gravitational softening.
}
\end{table}


\subsection{Cooling, Star Formation and Feedback}

{\sc ChaNGa} tracks the non-equilibrium abundances of H and He. In
addition to H and He cooling processes, we include cooling from metal
lines, metal diffusion \citep{shen10} and Compton cooling. A uniform,
time variable cosmic UV field from \cite{HM12} models photoionization
and heating. We assume that the cosmic UV field is generated in structures
above the WDM filtering scale, and it is then identical in the two
cosmologies. An approximate treatment of self-shielding is included in some runs
following \cite{pontzen10}: for each gas particle the strength
of the UV background used for ionization equilibrium and heating
calculations was reduced by the mean attenuation for particles of that
density. The SF and blastwave feedback recipes are similar to \citet{G10}. Briefly, star formation occurs stocastically when 
cold ($T<10^4$ K), virialized gas reaches a threshold density, and follows a 
Schmidt law:
\begin{equation}
{d\rho_*\over dt}=0.1 {\rho_{\rm gas}\over t_{\rm dyn}} \propto \rho_{\rm gas}^{1.5},
\label{eq:SFR}
\end{equation}
where $\rho_*$ and $\rho_{\rm gas}$ are the stellar and gas densities and $t_{\rm dyn}$ is the
local dynamical time. 
The force resolution of our simulation enables us to adopt a density
threshold for star formation of 100 atoms cm$^{-3}$, as the local
Jeans length at this density and a gas temperature of $T=10^3\,$K is
resolved with more than 6 SPH smoothing lengths.  A Kroupa
\citep{kroupa2001} IMF is assumed.  For consistency with our previous
works that reproduced the properties of galaxies over a range of
scales, supernova feedback is implemented according to the “blastwave”
feedback scheme \citep{stinson06}, where, in order to prevent
artificial cooling of gas particles heated by SNe, cooling is
temporarily shut off for typically a few million years. The result of
limiting star formation to dense gas regions is that feedback energy
is concentrated and, therefore, able to drive outflows
\citep{agertz14}.  In this work we adopted the same SF parameters
(density threshold 100 amu/cm$^3$, c$\star$~=~0.1 and 100\% of SN
energy coupled to the ISM) as in \cite{shen14}. We have verified that
this SF approach generates realistic SF efficiencies similar to
simulations where the SF efficiency is driven by the local H$_2$
fraction \citep{christensen14b}. The wind loading factors
generated by the 'blastwave' model are slightly lower than in a scheme
where the time scale for inefficient cooling is explicitly regulated
by a conduction subgrid model \citep{keller14} which
we plan to adopt in future works.

Each CDM and WDM halo was run several times, following a sequence of
increasing complexity in the description of ISM physics and SN
feedback that was identical for both cosmologies.  This sequence of
runs has the specific goal of differentiating the effects of changing
the DM 'scaffolding' vs the effects of using different implementations
of SF-related processes (see Table 1). Runs 'g1' include metal
cooling, cosmic UV and SN feedback, 'g3' include also HI shielding
from UV radiation, while 'g5' also include 'early' feedback from young
stars following \cite{stinson12}. The amount of energy from young
stars (often referred to as 'early feedback' that is injected into the
ISM as thermal feedback is 10$^{49}$ erg per solar mass of stars
formed. Cooling of the gas is not shutoff for 'early feedback'. The
total amount of energy injected into the ISM is then $\sim$ 2 $\times$
10$^{49}$/\Msun of stars formed, comparable to the energy coupled to
the ISM in other recent stellar feedback schemes
\citep{hopkins13,vogelsberger13,angles14}.  {\it This work focuses on
  the 'g5' runs}, as they include the most complete description of ISM
and feedback physics.  For reference, the SF and feedback of the `g5'
model resembles the one in \cite{dicintio14}, while `g1' resembles the
runs in \citep{shen14}. However, those previous runs do not include
the new SPH implementation.  We highlight the differences between the
different implementations at different points in the paper. We plan to
extend our analysis to a large sample of small systems in future work.

\begin{figure}
\includegraphics[width=80mm]{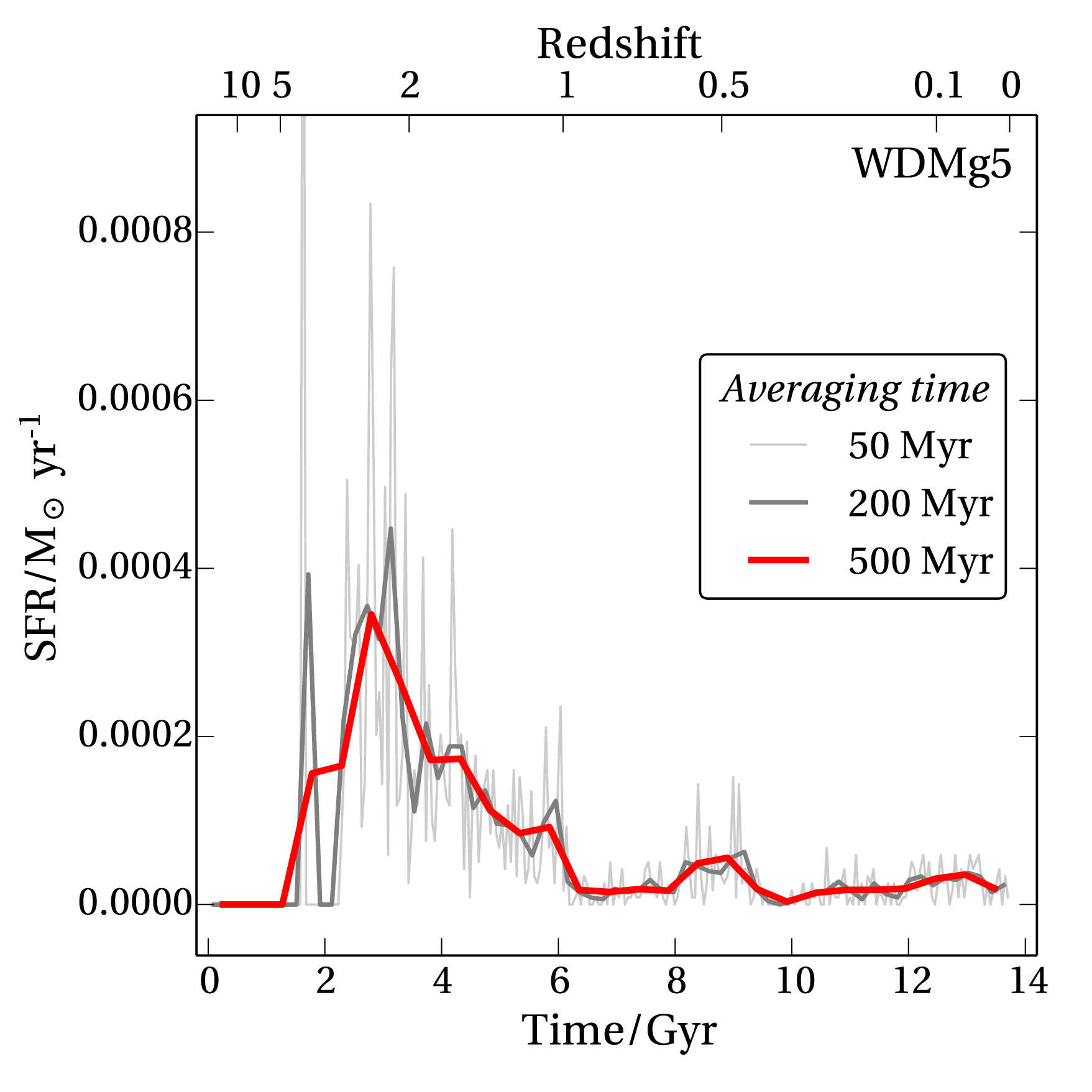}
\caption{ {\it The SFR of the galaxy in the WDM cosmology.}  SFR is
  measured including all the stars within the central 1kpc t z$=$0.
  The reference WDMg5 run has delayed SF compared to the CDM runs. The
  WDM run forms less stars overall. This delay is independent of the
  details of SF and feedback.  }
\label{fig3}
\end{figure}

\section{The Assembly of DM and stars in galaxies}

The main properties of the simulated halo are summarized in Tables 1
\& 2.  Halos were identified with the halo finder AMIGA
\citep{knollmann09}.  By z=0, the final halo mass is approximately the
same in CDM and WDM cosmologies, as the halo mass (10$^{10}$ \Msun) is
above the 2keV filtering scale. The main halo accumulated 50\% of its
final mass by z$\sim$ 2.5 in both cosmologies.  However, the assembly
histories are markedly different at higher z where the filtering of
small scales in the WDM model becomes important.  At z$=$4, 25\% of
the CDM halo final mass is in collapsed, star forming progenitor
haloes, while the main progenitor of the WDM halo contains only 7\% of
its final mass. This delay in halo assembly is typical of WDM
cosmologies implying that SF is absent or minimal during the first two
Gyrs (compare Fig.~\ref{fig2} with Fig.~\ref{fig3}). A delayed SF in
WDM models, especially in small systems, had been recently suggested
in a semi analytical model \citep{calura14}. Most of the baryons of the
WDM galaxy accreted onto the main halo from smooth accretion and not
through mergers with smaller halos.  The SFH of models including
shielding (g3 and g5) and early feedback from young stars (g5) are
extended, with several small bursts occurring over the whole Hubble
time.  the `CDMg1', which lacks both the above physical processes
shows a more intermittent, very bursty SF.

\begin{figure*}
\includegraphics[width=140mm]{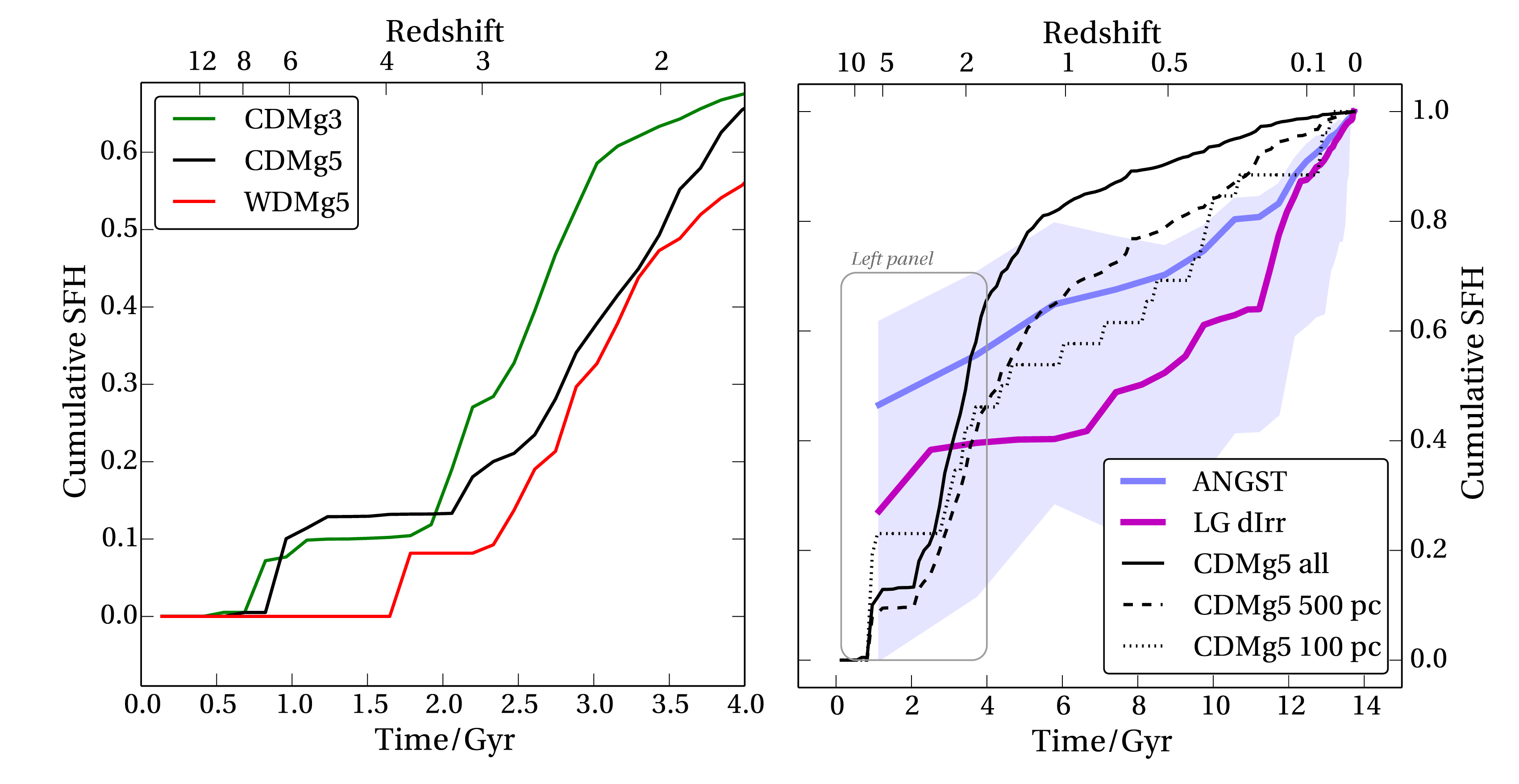}
\caption{ {\it Left: The cumulative SFH (i.e., the fraction of total
    stellar mass formed prior to a given epoch, normalized to one at
    the present) within the 500 pc of the simulated galaxy in the CDM
    and WDM cosmologies}. CDM: solid (g5), green (g3) lines, WDM (g5):
  red line.  Overall the SFH of the simulated galaxy reproduces the
  rapid rise and the subsequent linear growth of the ANGST sample (see
  right panel), but SF starts one Gyr later in the WDM model. {\it
    Right: The cumulative SFH as a function of aperture of our
    standard implementation CDMg5}.  The SFH from the simulation is
  measured including spherical regions of different radius, but all
  centered on the galaxy center (black: all, dashed: 500 pc, dotted:
  100~pc). ANGST average: blue, Local group: magenta.  The shaded area
  shows the dispersion of the ANGST sample.  The differences in the
  simulated SFHs illustrate how the center of the simulated galaxy is
  populated by younger stars while the outer regions consists mainly
  of older stars, likely scattered outward during the process of core
  formation. This radius vs age bias may explain the difference
  between the ANGST and the LG sample, the latter sample stars in the
  very central regions (55-300~pc) of relatively nearby systems.}
\label{fig4}
\end{figure*}

By z$=$0. `g5' galaxies in both cosmologies show
a rotationally flattened stellar component and an extended HI disc,
with no sign of a central bulge component. However the stellar disk is
quite thick and dynamically hot. Thick stellar disks are a
characteristic of feedback implementations that model radiation or
stellar winds from young stars \citep{stinson13,trujillo13,hopkins13}.
By z=0, the amount of stars formed in the WDM model is only 20-40\% of
the CDM counterpart, depending on the specific SF model adopted: `g3'
runs (where gas is shielded) form the largest amount of stars, while
`g5' (where feedback from young stars is also included) form the least.
While all CDM realizations show a cold gas/stars mass ratio around one
or larger when measured at z=0, WDM models are several times more HI
rich than observed dwarfs of similar mass, which typically have
HI/stellar mass ratios closer to unity
\citep{geha06,mcconnachie12}. We speculate that the longer
depletion time scale of cold gas in WDM could be due to a drastically
reduced interaction rate with dark subhalos, which could induce disk
instabilities \citep{chakrabarti13,maccioWDM}. Alternatively, this
could also be due to the later assembly of the WDM halo and the
delayed onset of gas cooling, which then settles in a larger, lower
density disk. We verified that the WDM halo accretes only one subhalo
with circular velocity V$_c$ $>$ 10 km/sec, at z$\sim$3\footnote{V$_c$
  is defined as sqrt(M/r) throughout the paper, with G$=$1}.

\section{SFH Histories and Color Magnitude diagrams}

In this work we compare the SFHs of our models with those of the
galaxies in two HST studies: the ANGST sample and of a subsample of
field Local Group dwarfs with deeper observations than ANGST
\citep{weisz14}.  The SFHs of the real galaxies were
reconstructed from the analysis of their resolved stellar populations
\citep{dolphin02,tolstoy09,mcquinn10}.  For the purpose of comparing
simulations with observations is important to note that the typical
region of each galaxy surveyed by HST varies with the galaxy distance.
Galaxies in the 'Local Group' sample have been observed to fainter
magnitudes, but only out to (typically) 100-300 pc. Stellar
populations in the ANGST sample have been typically observed out to
1-2kpc from the dwarf centre. This aperture effect may lead to
potential biases in the reconstruction of their SFHs. We discuss
effects of this aperture bias, and will explore it in the context of
simulated population gradients in future work (M.Teyssier et al. in
prep.).

\begin{figure*}
\includegraphics[width=110mm]{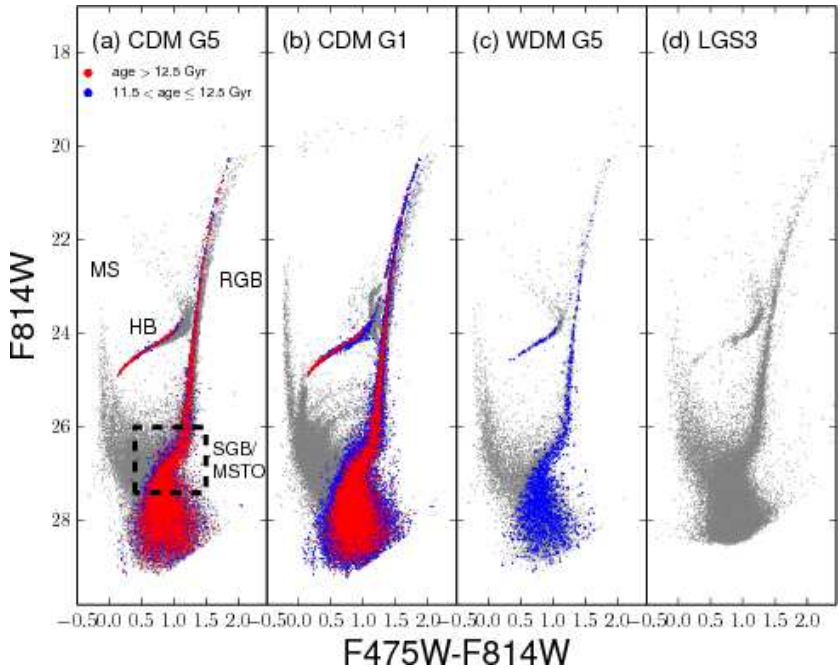}
\caption{ {\it Mock color-magnitude diagrams (CMDs) for select
    galaxies from our simulations.}  The CMDs have been designed to
  mimic deep HST observations of Local Group dwarf galaxies
  \citep{cole2007}.  We have highlighted features important for
  measuring the SFH of a galaxy (upper main sequence, MS; red giant
  branch, RGB; horizontal branch, HB; sub-giant branch, SGB; oldest
  main sequence turn-off, MSTO and color-coded stars that are between
  11.5 and 12.5 Gyr old (blue) and older than 12.5 Gyr (red). CDMg5
  (and CDMg3, not shown) has a CMD that is qualitatively similar to
  those observed in real local dwarfs such as LGS3 \citep{hidalgo11},
  which is shown in panel (d). In contrast WDMg5 is deficient in
  ancient stars, as it can be seen by the lack of of a blue horizontal
  branch compared to the LGS3 CMD (at a color of $\sim$ 0.5 and a
  magnitude of $\sim$ $\sim$ 24.7).  CDMg1 (the run with no
  self-shielding and early feedback) contains fewer, discrete bursts throughout its  lifetime, neither of which are usually observed in low mass galaxy
  SFHs.
}
\label{fig5}
\end{figure*}

Fig.~\ref{fig2} and Fig.~\ref{fig3} show the SFR of the CDM and WDM
halos as directly measured in the simulations at z$=$0. The SFHs
include only stars less than 1 kpc from the centre of the galaxy, to
be consistent with the typical aperture of galaxies in the ANGST
sample. Half of the mass in stars is located within a 1 kpc radius
from the center in the CDM galaxy, within 700pc in the smaller mass
WDM galaxy.  In CDM the first stars are formed as early as z$=$8,
while in WDM, star formation started one Gyr later, by z$\sim$4.  The
'g1' galaxies (lacking self-shielding) have episodic star formation,
while all other runs have bursty, but continuous SF. The 'g3' runs,
that include shielding, form about twice as many stars as our standard
run 'g5', that includes both self shielding and 'early feedback' from
young stars (Table 2).  Observations strongly support bursty SFHs in
small galaxies, with as many as 80\% stars being formed in bursts
\citep{kauffmann14} and a relatively high SF efficiency at high z
\citep{madau14b}.

Fig.~\ref{fig4} shows the cumulative SFHs of the CDM and WDM models against
those from the ANGST and Local group samples \citep{weisz14}.  The
average SFH of the ANGST and Local Group samples shows even older
populations than our simulated galaxy.  This difference likely depends
on the detailed assembly history of our galaxy. However, in all our
simulations the CDM galaxy forms about 10\% of their stellar mass
before the WDM galaxy even starts forming stars. This earlier relative
start of SF in the CDM models is independent of the SF model used, and
persists even when 'early' feedback is included (which overall delays
SF, \cite{stinson13}).  As soon as the WDM galaxy starts forming
stars it catches up rapidly and by z $\sim$ 1.5, 50\% of present day
stars have formed in both models.  SF after z = 2 then depends mostly
on the implementation of SF and feedback.  Fig~\ref{fig4} (right panel)
compares the cumulative SFH of our simulated galaxy CDMg5 compared to
the average SFHs of the ANGST Local Group (LG) samples.  Analysis of
the simulations allows us to understand the difference in the SFHs
between the two samples, with the 'LG sample' galaxies apparently
having much younger stellar populations. Part of the observed
difference can be reproduced by calculating the cumulative SFH of
stars selected inside smaller and smaller radii of the simulated
galaxy.  In CDMg5, only 10\% of halo stars (a subsample with minimum
distance 1~kpc from the galaxy center) are younger than 8 Gyrs,
compared to 40\% of the stars within a radius of 200~pc.  We verified
that at z=3.4, the mean radius of all stars in the main halo of CDMg5
is $<$r$>$=0.66 kpc, while taking those exact same stars but tracing
them forward to z=0, the mean radius is now $<$r$>$=1.22 kpc. This
finding shows how the same 'dynamical heating' process that removes DM
from the galaxy centers can also act on stars (an equally
collisionless component), resulting in older stars being moved outward
on less bound orbits \cite[see also][]{maxwell12}. {\it  Together these
results predict real dwarfs to have 1) a significant population of old
stars, 2)  age gradients, with younger stars being more
centrally concentrated. These result are independent of the DM and
SF models used in this work.}

\begin{figure}
\includegraphics[width=87mm]{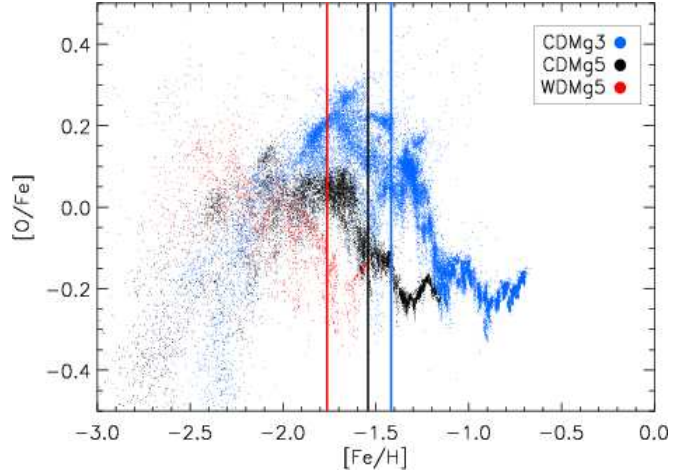}
\caption{ {\it The [O/Fe] vs [Fe/H] for CDMg3, CDMg5 and WDMg5.}  The
  abundance trend is set by the SFH rather than the DM model. As 'g3'
  is the most efficient star former, the ISM is enriched by of SNII
  (high O/Fe) before the SNIa kick in, and overall it reaches the
  highest [Fe/H].  The g5 runs form less stars, and the initially
  lower SFR lead to lower O/Fe at low metallicities.  The "knee" kicks
  in at lower [Fe/H] for lower SFR, and the overall highest [Fe/H]
  reached is just dependent on the total stellar mass formed.  The
  vertical lines show the average metallicity for dwarf galaxies with
  stellar masses equal to the simulated ones, from Kirby et
  al. (2014).  }
\label{fig6}
\end{figure}

In Fig.~\ref{fig5} we show the artificial CMDs obtained from the stellar
populations of different simulations in our sample (see Appendix for
details). These CMDs include typical observational effects such as
resolution and limiting magnitude of HST observations of Leo A, a
typical nearby dwarf $\sim$ 800kpc away: with a CMD completeness limit
of $\sim$ 29.0 in F475W and 27.8 in F814W \citep{cole2007}. These CMDs
qualitatively show the features in the SFH described above. The lack
of old stars in the WDMg5 run and the very episodic SFH of CDMg1 (the
CDM run with no self-shielding) can be clearly identified and are
not features commonly observed in the CMDs of nearby dwarf
galaxies. Artificial CMDs such as these will be used in future
comparisons between observational and simulated samples.

\begin{figure*}
\centering{
\includegraphics[angle=0,width=80mm]{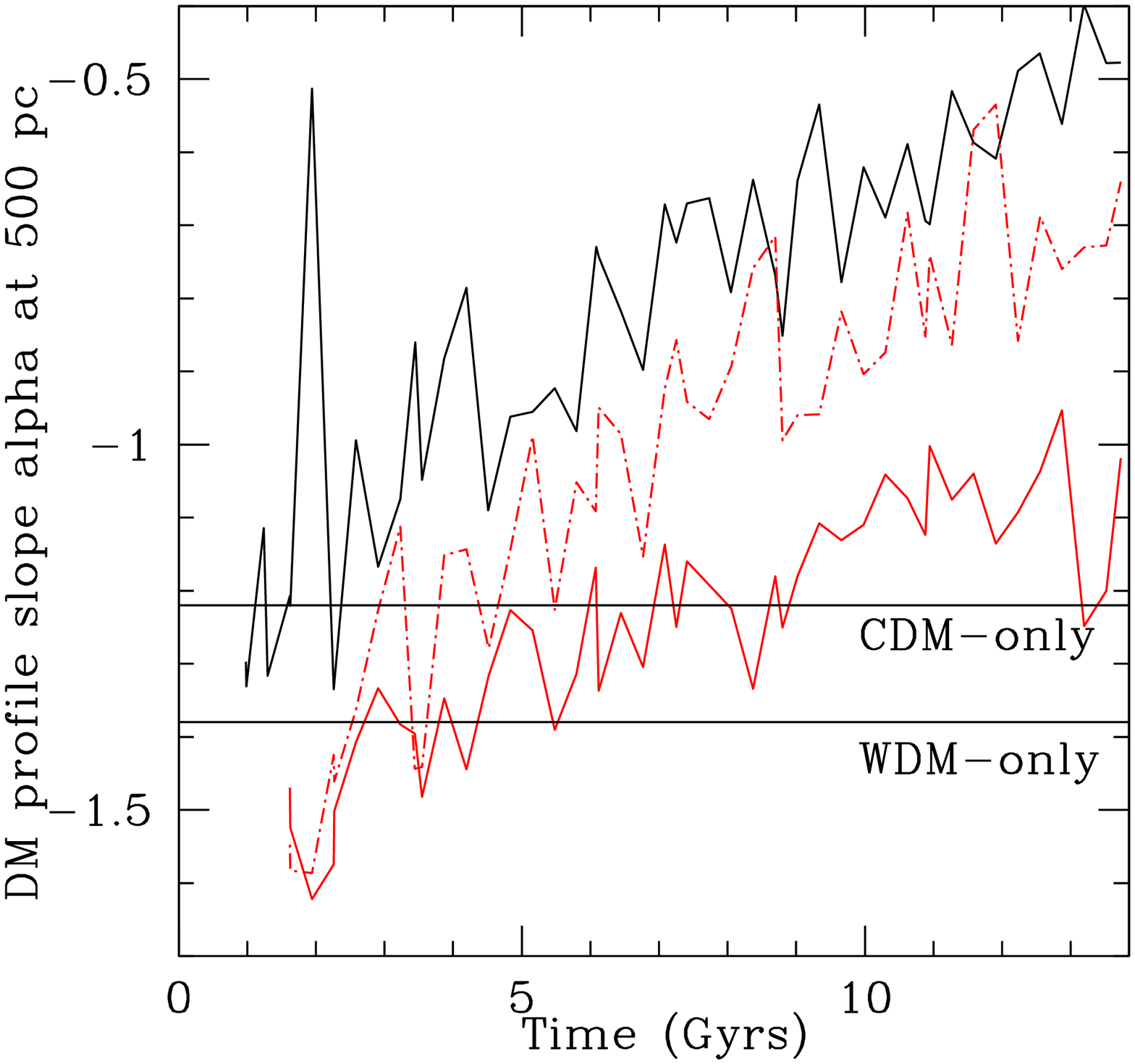}
\raisebox{-0ex}{\includegraphics[width=80mm]{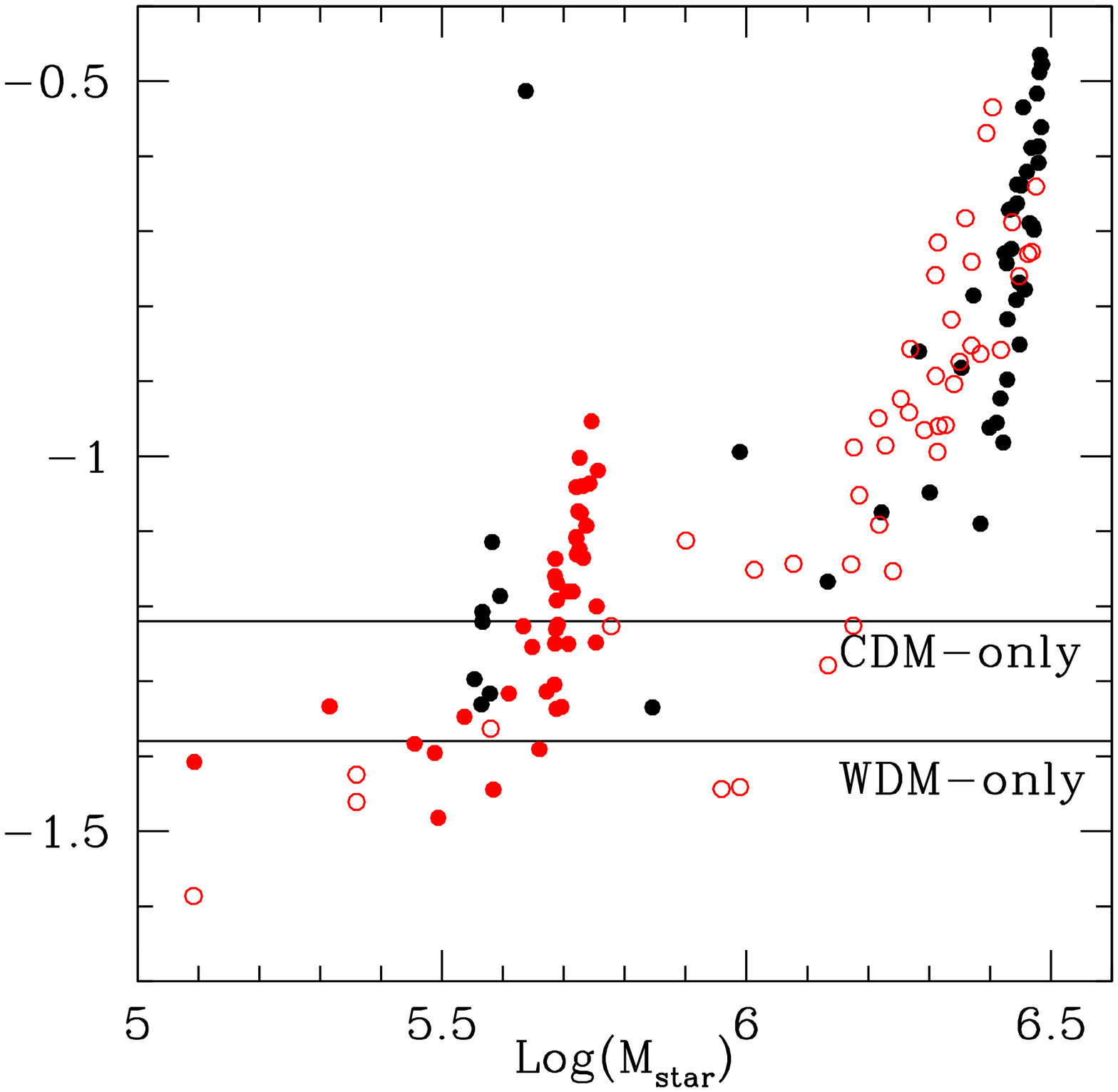}}
}
\caption{ {\it Left: The time evolution of the slope $\alpha$ of the
    CDMg5 (solid black) and WDMg5 (solid red) halos, measured at 500~pc.}  The red
  dash-dotted line shows the WDMg3 run, which has a higher SF
  efficiency, leading to a final flatter DM profile. These results
  highlight how DM 'heating' is a continuous process which begins at
  high-z and depends strongly on the SF efficiency, (higher
  efficiency leading to flatter profiles) rather than on the details of the
  CDM vs WDM cosmologies. As a reference, the DM $\alpha$ slopes at
  z$=$0 of the DM-only simulations are shown as horizontal lines.
  {\it Right: The slope of the DM profile measured at 500 pc vs. the
    galaxy stellar mass} for several time outputs of the main
  progenitor of CDMg5: solid dots, WDMg5 (solid red) and WDMg3 (red
  circles). Dashed horizontal lines show the z$=$0 of the equivalent
  DM-only runs.}
\label{fig7}
\end{figure*}

In Fig.~\ref{fig6} we examine metallicity trends in the CDMg3, CDMg5, and WDMg5.
We track oxygen as a proxy for $\alpha$ abundances.  Overall, the
chemical trends are driven by the SFH of each simulation.  CDMg3 is
the simulation with the most efficient star formation.  With a higher
early SFR then the g5 models (twice as many stars formed in the first
3 Gyrs), CDMg3 is initially dominated by enrichment from SN\,II, with
a high [O/Fe] ratio.  At later times, SN\,Ia begin to contribute
predominantly Fe, causing the [O/Fe] to drop as [Fe/H] increases.
Because it forms the most stars overall, CDMg3 reaches the highest [Fe/H]
of any of the models. The g5 runs, on the other hand, have such low
SFRs that SN\,Ia are able to contribute at lower [Fe/H].  The `knee'
in these plots occurs at lower [Fe/H] for a lower SFR, and the runs
that form the most stars are able to reach a higher [Fe/H] ultimately.
The vertical lines show the mean [Fe/H] of field dwarfs of similar
stellar mass as their color coding from Kirby et al. (2014). While a
direct comparison would require a more detailed comparison (M.Teyssier
et al. in prep.) they show that the average metal content of the
simulated galaxies is consistent with observations.

\section{Mass distribution: evolution and comparison with observations}

Fig.~\ref{fig7} (left panel) shows the evolution of $\alpha$ (here
$\alpha$ is the DM density profile slope measured at 500~pc), for the
main progenitor of various models as function of time. This plot
emphasizes how the formation of cores is a continuous process driven
by prolonged and bursty SF. Core formation happens later in models
where halo collapse (as in WDM), or 'young stars' feedback, delays the
main onset of SF.  In Fig.~\ref{fig7}. (right panel) we show instead
the dependence of $\alpha$ on the amount of stars formed in a few
runs. Data points belong to the different progenitors of a given z$=$0
galaxy, so this is equivalent to a time sequence \cite[see
also][]{madau14a}.  The horizontal lines mark the slope of the DM
profile in the DM-only runs. This plot shows how the 'DM heating'
process is important even when only 10$^6$ \Msun of stars have formed,
independent of the details of SF and feedback implementations, as long
as they create rapid outflows. This results largely confirm those of
previous works \citep{teyssier13,PG12,madau14a} in more
massive halos.  In all simulations cores, once formed, persisted to
  the present time.

\begin{figure}
\includegraphics[width=80mm]{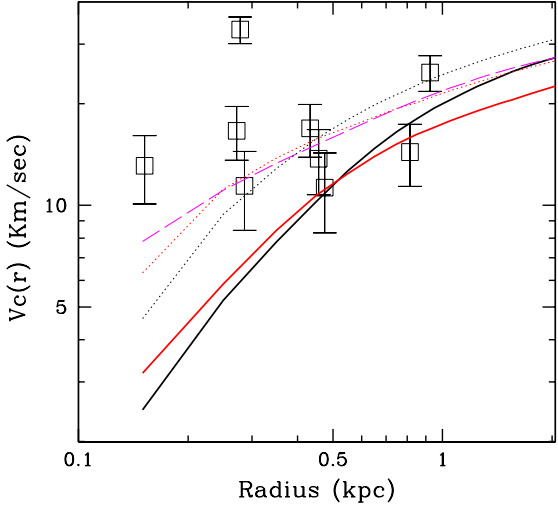}
\caption{ {\it The rotation curve (defined as V$_c$$=$sqrt(M/R), with
    G$=$1, including both DM and baryonic matter) of the simulated
    galaxies vs. the observational estimates of field LG dwarfs with
    stellar masses in the 10$^6$-10$^7$ M$\odot$ range.} Black: CDMg5,
  Red: WDMg5. Magenta dashed: CDMg1. Black and red dotted: CDM-only
  and WDM-only runs. Estimates from real galaxies are based on stellar
  kinematics. Error bars show a representative 5km/sec uncertainty
  rather than published errors. DM and stellar dynamical 'heating' due
  SF and feedback processes reduce the overall central density,
  irrespective of the DM component. As a result the amount of mass
  within the central 500pc (a scale well resolved in these
  simulations) is lowered. This process will help reconcile
    prediction of the mass distribution in dwarfs with the current
    observational constraints from Local Group galaxies}
    \citep{wolf10,weisz14,brook14}.
\label{fig8}
\end{figure}

In Fig.~\ref{fig8} we compare the total mass distribution (DM and
baryons) of our simulated CDM and WDM dwarf with the observational
constraints of a subsample of Local Group field galaxies
\citep[from][]{wolf10,mcconnachie12,weisz14}. 

The main result is that DM heating lowers the central total
  matter distribution of field dwarf galaxies, in both CDM and WDM
  cosmologies. By the present time the central mass distribution of
  all the runs with SF is remarkably similar in both cosmologies, with
  circular velocities at 0.5 kpc of the order of V$_c$= sqrt(M/r)
  $\sim$ 10 km/sec, consistent with estimates of local field dwarfs of
  similar stellar mass.  As DM-only simulations often show an excess
  of central mass in systems with low internal velocities, the
  introduction of feedback would help bring theoretical predictions in
  agreement with current kinematic constraints, independently of DM
  being 'Warm' or 'Cold'. A larger set of galaxies is however required
  to make more quantitative statements.

Overall these different results show that the mass distribution of dark matter
and stars depend more on the efficiency of star formation rather than
on the type of DM (CDM vs WDM) or the slope of the primordial power
spectrum of density perturbation at small scales. Models with more
efficient SF create more DM and stellar 'heating', erasing difference
driven by the details of initial power spectrum, assembly history and
formation times of the parent halos, which affect the DM halo
concentration \citep{gk14}. In following work we will study the
number density and observed circular velocities of a large sample of simulated
field dwarfs (Brooks \& Papastergis in prep.) and compare the DM profiles of CDM
and SIDM halos (Bastidas--Fry et al 2014, in prep.).

\section{Conclusions} 

We have simulated to z$=$0 the evolution of a small (halo mass $\sim$
10$^{10}$\Msun) field dwarf galaxy in $\Lambda$CDM and $\Lambda$WDM
cosmologies. Our goal was to identify differences in the observable
properties of the galaxy and its DM distribution once baryon processes
are included together with non-standard DM. These simulations have a
nominal force resolution of 60pc and gas forces resolved down to
$\sim$~10pc. For both cosmologies, we performed a series of runs where
we modeled the effects on SF of cold gas shielding and the deposition
of energy from young stars in the ISM. The SPH implementation better
follows Kelvin-Helmoltz instabilities and gas interactions at the
cold/hot interface.

\begin{itemize}

\item Due to the later assembly and suppression of substructure in the
  the WDM model, the WDM and CDM galaxies differ significantly in some
  properties of their stellar components. In the WDM model, SF begins
  1.5 Gyrs later, is more inefficient and results in a galaxy that has
  a higher and somewhat unrealistic HI/stellar mass ratio (closer to
  unity in real galaxies and in the CDM realizations). These
  differences are independent of the details of SF and feedback.  The
  effect of less efficient SF and increased HI mass in WDM models may
  have a strong effect on the luminosity and velocity
  distribution function of field galaxies and needs to be studied in
  larger simulated samples.

\item The inclusion of explicit `baryonic physics' is the major driver
  of the central distribution of galaxies.  Rather than being set by
  intrinsic differences between the Cold and Warm DM models, the
  matter distribution at the center of field dwarfs is largely set by
  the amount of stars formed \citep[with more stars forming resulting
  in the 'heating' of larger amounts of DM and
  stars][]{maxwell12,dicintio13}. By the present time, both CDM and
  WDM galaxies show a reduced central density and flatter DM profile
  compared to their DM-only realizations, making the simulated
  galaxies consistent with the constraints set by kinematic estimates
  of the mass content of field dwarfs of similar stellar mass.  At
  this high resolution, DM removal within the central 500 pc becomes
  significant even as few as 10$^6$ M$\odot$ stars formed.

\item Feedback lowers the central mass content of
  10$^{10}$\Msun halos, bringing them into agreement with observations
  even when SF is relatively inefficient when compared to standard
  Abundance Matching models \citep[M$_{star}$/M$_{halo}$ $\sim$
  10$^{-4}$, see][]{munshi13,behroozi13}.  This result shows that the
  complexity of baryonic processes needs to be explicitly included in
  models of galaxy formation when comparing the prediction of CDM or
  other models with the combined constraints from the kinematic and
  number density of local field dwarfs \citep{klypin14,brooks14}.

\begin{figure*}
\includegraphics[width=130mm]{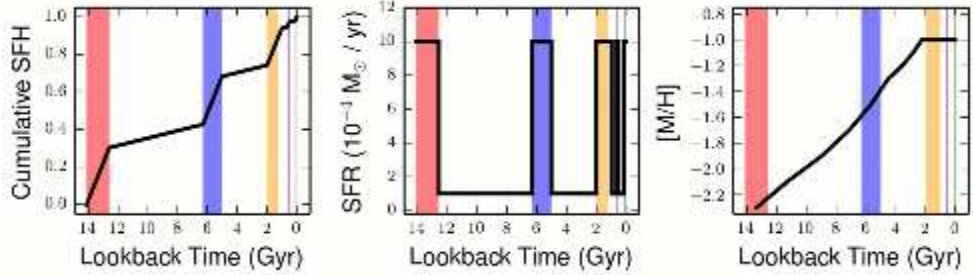}
\caption{{\it Appendix A: The mock star formation history (SFH) and
  age-metallicity relationship (AMR) used to constructed the schematic
  CMDs in Fig.~\ref{fig10}}  The left panel shows the cumulative SFH, i.e., the
  fraction of stellar mass formed prior to a given epoch; the center
  panel shows the absolute SFH; and the right panel shows the
  AMR. In this model 80\% of the stars is formed
in bursts. Note that the x-axis shows 'lookback time' and not the time
  since the Big Bang as in previous plots.  In each panel, we have
  highlighted each burst in color: 0-0.1 Gyr (green), 0.5-0.6 Gyr
  (purple), 1-2 Gyr (orange), 7-8 Gyr (blue), 12.5-14 Gyr (red).  The
  SFH is modeled after a bursty dwarf galaxy with quasi-periodic
  bursts superimposed on a low-level constant SFR. 
}
\label{fig9} 
\end{figure*}

\item A qualitative comparison between artificial CMDs and those
  obtained with HST on samples of local galaxies favors dwarf galaxies
  formed in a CDM cosmology, with very old stellar populations and a
  continuous SF comprised of several small bursts when properly time
  resolved. When plotted with a low time resolution, the SFH of dwarf
  galaxies appears smoother, being the superposition of timely and
  spatially independent SF events. In all galaxies, the older stellar
  populations are less radially concentrated, possibly the effect of
  energy transfer not just to the DM but also to stars, going a long
  way resolving the apparent discrepancy between the average SFH
  recovered from the full ANGST sample (with typical apertures of 1
  kpc) and those of the closest field dwarfs in our Local Group (where
  the limited HST field covered a much smaller central area, typically
  of 50-150pc in radius).

\end{itemize}

This is one of the first works to compare the {\it observable}
properties of the resolved stellar populations of faint field dwarfs
with the predictions of CDM and non-standard DM models. Given the
sample of just one halo the results are necessarily preliminary.  In
future work we plan to extend this study to a larger sample, allowing
us to predict the number density of field dwarfs \cite{klypin14}.  At
high redshift, a DM scenario with reduced small scales power
\citep[see also][]{gk14b} may leave a clear signature in the abundance
and SF rates of faint galaxies \citep[stellar mass 10$^{7-9}$
\Msun][]{atek14}. In higher redshift systems, the delay in SF in WDM
models would be evident in the SFR/stellar mass relation
\citep[e.g.][]{noeske07,puech14,calura14}.  These measurements are now
becoming feasible through deep observations of field magnified by
lensing clusters \citep{menci12, pacucci13,alavi14,ferrara14}.  Future
hydrodynamical simulations will help disentangle the effects of SN
feedback from that of the underlying DM or dark energy models,
provide a robust counterpart to the stronger constraints coming from
studies of the properties of small field galaxies, and better guide
predictions for DM direct detection experiments \cite[see review
by][]{brooks14r}.

\medskip
\section*{Acknowledgments}
FG and TQ were funded by NSF grant AST-0908499.  FG acknowledges
support from NSF grant AST-0607819 and NASA ATP NNX08AG84G. Support
for DRW is provided by NASA through Hubble Fellowship grants
HST-HF-51331.01, awarded by the Space Telescope Science
Institute. Simulations were run on Pleiades (NASA HEC) and Blue Waters
(XSEDE) supercomputers. ChaNGa was developed with support from
National Science Foundation ITR grant PHY-0205413 to the University of
Washington, and NSF ITR grant NSF-0205611 to the University of
Illinois. Some images were created with the analysis and graphic
package Pynbody \citep{pynbody}. SRL acknowledges support from the
Michigan Society of Fellows.

\bibliography{bibref}

\bibliographystyle{mn2e}

\begin{figure*}
\includegraphics[width=130mm]{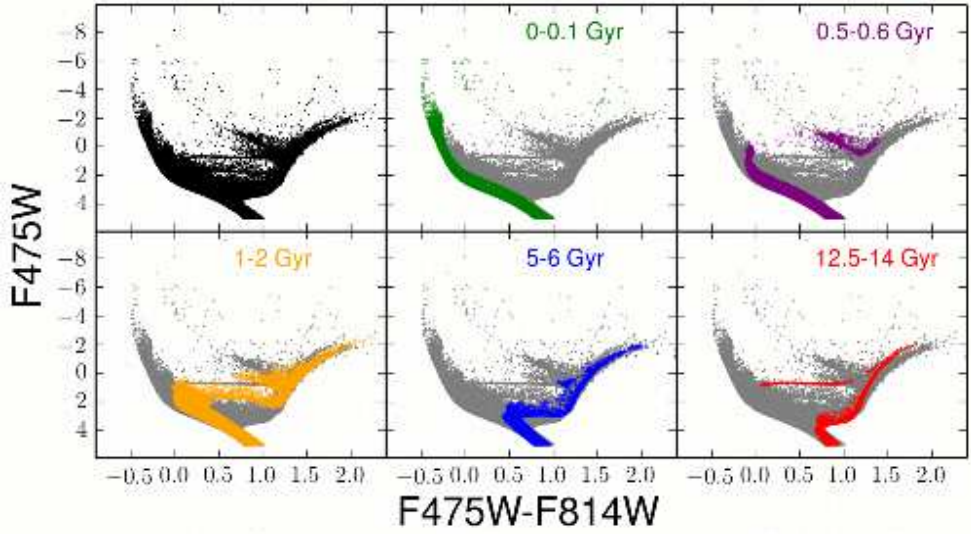}
\caption{{\it Appendix A: Simulated CMDs based on the SFHs from Fig.~\ref{fig9},} and observational uncertainties designed to mimic an HST observation of the LMC.   In each panel we have highlighted the bursts of star formation from Fig.~\ref{fig9} in an identical color.
}
\label{fig10} 
\end{figure*}

\section{Appendix- Features in the Color Magnitude Diagram}

In this section we provide a few examples of how the features of
artificial Color Magnitude Diagrams are able to provide useful
constraints on the star formation  and assembly histories of simulated
galaxies when compared with observations. This  discussion is
aimed at the non CMD specialist.

The resolved star color-magnitude diagram (CMD) of a galaxy can be
thought of as a linear combination of simple stellar populations
(SSPs), i.e., populations of a single age and metallicity, such as
star clusters.  The goal of CMD analysis is therefore to determine the
stellar mass and metallicity of each SSP that makes up the total CMD
of a galaxy.  Over the past 25 years, there have been over two dozen
algorithms aimed at reconstructing the SFH of a galaxy based on CMD
analysis, with comparisons of various techniques generally yielding
consistent solutions \citep{skillman02,monelli10a,monelli10b}.  Despite
the reliability of the general methodology, there remain several
outstanding challenges in accurately and precisely measuring a SFH
from a CMD.  In this section, we qualitatively discuss the age
sensitive features available on an optical CMD, and discuss both
intrinsic limitations (e.g., age-metallicity degeneracies) and other
factors (e.g., CMD quality, uncertainties in stellar physics) that
affect the accuracy and precision of SFH determination even in high
quality data.

To help illustrate the connection between SFHs and CMDs, we make use
of simulated CMDs, which allow us to clearly highlight key
age-sensitive CMD features.  To construct these mock CMDs, we used the
SFH and age-metallicity relationship (AMR) shown in Fig.~\ref{fig9}, as input
into the CMD simulation tool from the CMD analysis software suite
\texttt{MATCH} \citep{dolphin02} along with the Padova stellar
evolution libraries \citep{girardi10}.  The SFH and AMR are designed
to mimic a bursty dwarf galaxy with bursts of SFR $=$ 10$^{-3}$
M$_{\odot}$/yr superimposed at select times onto a constant SFR $=$
10$^{-4}$ M$_{\odot}$/yr. The simulated CMDs in Fig.~\ref{fig10} have also been
convolved with observational uncertainties (e.g., photometric errors,
color and magnitude biases) that are designed to mimic an HST
observation of the LMC.  The CMDs are shown in the F475W (Sloan
g-band) and F814W (I-band) filter combination, which is commonly used
for resolved stellar population studies of nearby dwarf galaxies
\citep{cole2007,skillman02,ANGST09,monelli10a,monelli10b,hidalgo10,weisz14}.
The simulated CMDs and SFH share a common color-coding scheme,
allowing for a simple translation from bursts of star formation to
stars on the CMD.

Beginning at recent times, we see that stars from the youngest burst
(0-0.1 Gyr; green) can be found on the main sequence (MS), as well as
the blue and red core helium burning sequences (BHeBs, RHeBs) -- the
post-MS phase of evolution for stars more massive than a couple of
solar masses.  The age leverage for such young populations typically
comes from both the luminosity of main sequence turnoff magnitude
(MSTO; of M$_{F475W}$ $<$ $-$3 for an age of 0.1 Gyr) and the
luminosity function of the BHeB and RHeB stars, which are follow a
nearly monotonic relationship with age \citep{dohm-palmer97}.
While BHeBs and RHeBs are excellent age indicators in theory, they are
also rapidly evolving evolutionary phases whose physics is less
certain.  Uncertainties in the physical parameters (e.g., mass loss,
convective overshooting) make it challenging to accurately assign
absolute ages based on evolved stars alone. Instead, the MSTO provides
a more secure absolute age indictor, as the models of MS stars and
their evolution at the turnoff is generally better understood \citep{gallart07}.

The next oldest burst (0.5-0.6 Gyr; purple) has a MSTO that is less
luminous than the 0-0.1 Gyr old population.  Stars of this age also
occupy several distinct areas on the CMD including the faint end of
the BHeB and RHeB sequences, the thermally pulsating asymptotic giant
branch (TP-AGB) phase, the red clump (RC; m$_{F475W}$ $\sim$ 1.2,
m$_{F475W}$-m$_{F814W}$ $\sim$ 0).  Unlike the MSTO and BHeB/RHeB, the
TP-AGB and RC can be composed of stars of many difference ages and
metallicities, as we discuss below.  Thus, precise age determination
from these features alone is difficult.  Additionally, the physics of
these evolved phases is still not fully understood, which further
complicates our reliance on them for SFH reconstruction.  Typically,
the combination of the MSTO and the evolved phases (e.g., RC) yield
the best constraints on the CMD, although there is still not consensus
on this approach, and some prefer to only model the MSTO.

Increasingly older bursts (1-2 Gyr, orange; 5-6 Gyr, blue; 12.5-14
Gyr, red) show several common features: MSTO that decreases in
luminosity with increasing lookback age, occupation of the RC, and
stars that populate the red giant branch (RGB).  Star-formation that
occurred more than $\sim$ 1 Gyr ago no longer has BHeB or RHeB stars,
which have died.  In this age range, only the youngest population (1-2
Gyr; orange) produces luminous AGB stars, while only star-formation
that is at least several Gyr old produce a horizontal branch (HB).
The HB is known to be sensitive to age and metallicity, with a blue HB
forming for the oldest and most metal poor populations.  However,
uncertainties in HB models compromise its reliability as an accurate
age indicator.  Optimal SFH recovery comes from a CMD that extends
below the oldest MSTO.  For ancient SFHs, such deep CMDs contain both
the oldest MSTO and the sub-giant branch (SGB), whose colors and
luminosities allow for the separation of sequences as closely spaced
as $\sim$ 1 Gyr at the oldest ages possible.  The loci of such ancient
sequences are typically separated by 0.05-0.1 dex in color and
magnitude in the F475W and F814W filters.  The amount of separation
between the sequence is less for filters that are not as far apart in
effective wavelength (e.g., R and I), and higher signal-to-noise
ratios (i.e., longer integration times) are needed to achieve the same
precision on the ancient SFHs.  For younger ages, such deep CMDs are
guaranteed to contain the maximum amount of information possible for
SFH reconstruction.

Of course, not all CMDs extend below the oldest MSTO.  Due to stellar
crowding effects, beyond $\sim$ 300 kpc, the oldest MSTO of a galaxy
is no longer resolvable from ground-based telescopes.  Similarly,
beyond $\sim$ 1 Mpc, even HST-based CMDs are restricted to magnitudes
brighter than the oldest MSTO.  As a result, the majority of SFHs in
the literature have been derived from CMDs that do not contain the
oldest MSTO.  The net effect is that the resulting SFHs are limited to
age resolutions of several Gyr for ages larger than a few Gyr.  The
coarse age resolution of these SFHs is due to intrinsic limits of
shallow CMDs.  Well-known degenerates among evolved stars (e.g., the
age-metallicity degeneracy on the RGB; Cole et al.\ 2005) coupled with
reliance on the less-certain physics of these phases of evolution that
are typically overcome by the oldest MSTO, dominate the error budget
for SFHs from CMDs that do not contain the oldest MSTO \citep{dolphin02,gallart07}.

\bsp

\label{lastpage}

\end{document}